\documentclass[12pt,amsmath,amssymb,aps,floatfix]{revtex4-1}

\usepackage{graphicx}
\usepackage{bm}
\usepackage{color} 

\begin{document}

\title{Surface impacts and collisions of particle-laden nanodrops}

\author{Joel Koplik}
\email{jkoplik@ccny.cuny.edu}

\affiliation{Benjamin Levich Institute and Department of Physics \\
City College of the City University of New York, New York, NY 10031\\ }

\date{\today}

\begin{abstract} 
The surface impact and collisions of particle-laden nanodrops
are studied using molecular dynamics computer simulations.  The
drops are composed of Lennard-Jones dimers and the particles are
rigid spherical sections of a cubic lattice, with radii about 11
nm and 0.6 nm, respectively.  Uniform suspensions of 21\% and 42\%
particle concentrations and particle-coated drops 
are studied, and their behavior is compared to that of pure fluid
drops of the same size.  The relative velocities studied span the
transition to splashing, and both wetting/miscible and non-wetting/immiscible
cases are considered.  Impacts normal to the surface and head-on collisions 
are studied and compared.  In surface impact, the behavior of low-density
suspensions and liquid marble drops is qualitatively similar to
that of pure liquid, while the concentrated drops are solid-like on
impact.  Collisions produce a splash only at velocities significantly
higher than in impact, but the resulting drop morphology shows a
similar dependence on solid concentration as in impact.  In all
cases the collision or impact produces a strong local enhancement
in the kinetic energy density and temperature but not in the particle
or potential energy densities. Mixing of the two colliding species
is not enhanced by collisions, unless the velocity is so high as
to cause drop disintegration.
\end{abstract}

\maketitle

\section{Introduction} \label{intro}

The collisions and surface impacts of liquid drops have been studied
systematically for over a century \cite{worth,rein,yariv,lesser,marengo}, 
with motivations ranging from
the esthetics of a splash to understanding of the underlying fluid
mechanics to applications such as printing and sound generation in
rainfall.  More recently interest has expanded to drops of complex liquids
\cite{crooks,german,guemas,luu,moon}, and in
particular to drops containing solid particles
\cite{aussillous,nicolas,jaeger1,marston,jaeger2}, which is the focus
here.  In recent papers we have used molecular dynamics (MD)
simulations \cite{p1,p2} to study the impact of liquid drops on surfaces of 
different wettabilities, textures and shapes.  Such calculations are
naturally restricted to fairly small drops and limited time intervals,
tens of nanometers and nanoseconds, respectively, but remain relevant for a
number of reasons:
\begin{itemize}
\item MD results are controlled by the choice of atomistic interactions 
alone, independent of other assumptions.
\item As shown in \cite{p2}, in most cases the MD results are consistent with
experiment and continuum calculations using straightforward modeling
choices.  The principal exceptions occur when the surface attraction is
weak, where MD is perhaps more reliable anyway, and at higher Reynolds and
Weber numbers, where MD velocities are so high as to cause disintegration.
\item Nanodrops have their own potential applications, to heat transfer in 
micro- and nano-fluidic systems.  More generally the latter field is so new 
that is is worthwhile to explore how its small characteristic length and
time scales affect familiar continuum phenomena.
\end{itemize}

As a direct continuation of our previous work we begin by asking how the
presence of suspended solid particles alter the qualitative behavior of an 
impact with a planar solid surface.  Three types of drop are considered: 
a low-concentration uniform suspension, a higher-concentration suspension
and a liquid marble consisting of a drop decorated by solid particles on
its boundary.  The solid particles are rigid spherical sections of a cubic
lattice, and the fluid consists of Lennard-Jones dimers. In the first case, 
we present results for a suspension with 21\% solids concentration, where 
the percentage refers to the fraction of solid atoms.  This case is found
to have the qualitative behavior as lower concentrations since, as we shall 
see, the particles simply advect with the fluid during the impact.
Significantly higher concentrations of solid do change the behavior in
impact, and we discuss a 42\% solid concentration drop as an example.  Here
the aim is to present a comparison and an illustration of different
behavior, but we do {\em not} claim that the specific results hold for all 
high concentrations nor for all choices of the solid-liquid interaction.
The third system considered here is a liquid marble 
in which the particle reside on the surface of a drop.  In MD
calculations the difference between the marble and suspension drops lies 
in the strength of the attraction between solid and liquid atoms, as
discussed below.  Many other cases are possible (and equally amenable to
calculation) since drop and particle sizes, interaction form and strength,
temperature, surface characteristics, etc., can all be varied, but we have
limited this paper to three hopefully representative cases.

After discussing impact on solid surfaces, we turn to the head-on {\em
collisions} of liquid drops and the same three suspension drops as above. 
Drop collisions constitute a well-studied and significant subject in
itself, with numerous laboratory experiments and an extensive modeling
literature; see, for example, \cite{ashgriz,qian,eggers,roisman,paulsen}. 
More recently, an MD simulation literature on collisions has begun to appear 
\cite{greenspan,koplik,svanberg,zhao,juang,pothier}
as well, but here our principal motivation is to compare the
two processes and elucidate the effects of particulate content on the
interactions between deformable objects and with rigid surfaces.
(An earlier paper \cite{kompinski} makes a similar comparison for the case of
droplets residing on a solid surface, but the focus is somewhat different.)
In addition, we have the practical motivation that an MD impact simulation 
is very easily converted to a
collision simulation by simply removing the solid surface and duplicating
the drop, so the computation and analysis are nearly identical.  Once again 
the number of possible variations in collision
parameters is immense and here we focus on the same three comparison cases 
only.

The outcome of a surface collision naturally depends on the wettability of the
liquid and solid involved, and here we consider the extreme cases of 
complete wettability (strong attraction between solid and liquid atoms)
and non-wettability (no attraction and just a short distance repulsion).
The wettability of the solid particles inside the drop matters as well, but 
we wish to consider suspensions that are stable in equilibrium, which
requires an attractive interaction.  Likewise, collisions between drops of
the same material are not the same as heterogeneous collisions, so we also
consider the contrasting case of collisions between drops of two immiscible 
liquids.  

Some aspects of the outcome of a surface impact or collision are evident in
snapshots of the intermediate and final states, but further
understanding may be obtained from the internal density, temperature,
pressure and energy density fields. In an MD simulation, any field
expressible in terms of atomic variables may be calculated, and we do 
so here in order to understand how a drop's initial translational kinetic 
energy is either distributed within itself or lost to the exterior.
Two related issues are the coherence of the drop -- the extent to which the
solid particles and liquid stay together during and after impact, and  
the degree of mixing between the contents of 
the two colliding species and in particular the degree to
which it is enhanced by a hard contact.  These question are discussed below
in terms of the time-dependent species density fields.

The organization of the paper is as follow.  A description of the MD
procedure, which very closely resembles that in the previous papers
\cite{p1,p2}, 
in given next, followed by
two sections describing the results of impact and collision simulations, 
respectively, focusing on a qualitative comparison of shape change.  Next, 
we present a more quantitative analysis of the energy transfer where we
analyze the local potential and kinetic energy and temperature fields as a
function of time.  A similar study of the density fields for the individual 
atomic species is then the basis for a discussion of the degree of mixing 
resulting from drop impact, and is followed by our conclusions. 

\section{MD methods} \label{md}

We are interested in spherical drops of a generic uncharged Newtonian 
liquid containing solid spherical particles distributed uniformly throughout 
the interior (suspensions) or on the surface (marbles).  
Since the drops are nanometer sized and subject to strong thermal fluctuations 
the adjectives ``spherical'' and ``uniform'' are necessarily approximate.  
The liquid is composed of Lennard-Jones dimers, which have the potential
energy functions
\begin{equation}
V_{LJ}(r)=4\epsilon\left[\left({\sigma\over r}\right)^6-c\left({\sigma\over
r}\right)^{12}\right] +\delta V(r)\qquad\qquad
V_{FENE}=-{1\over 2}kr_0^2\log{(1-r^2/r_0^2)}
\label{eq:ljf}
\end{equation}
The Lennard-Jones (LJ) potential $V_{LJ}$ acts between all atoms within a 
cutoff distance $r_c=2.5\sigma$, and the potential is shifted by a linear 
term $\delta V$ so that the force vanishes at the cutoff.  The constant $c$
in $V_{LJ}$ adjusts the strength of the attractive interaction; its value 
is unity for fluid atoms of the same species and for fluid-solid
interactions for a liquid  
which completely wets a solid surface, but is zero for the
interaction between atoms of ``immiscible'' drops and for the fluid-solid
interactions of a completely non-wetting liquid.  The fluid-particle atomic
interaction
coefficient is again unity for a suspension, where the fluid completely wets
the particles which then prefer the interior of the drop, but we use a
weaker value $c=0.75$ for the marble case so that the particles are less
attracted and prefer the drop surface.  This specific numerical value was
chosen because it produces a 90$^\circ$ contact angle for a drop on a planar
solid surface.   The FENE (finitely extensible linear
elastic) potential acts between each pair of atoms in a dimer molecule.
Dimers are used in place of monomers because the monatomic Lennard-Jones
system has a very high vapor pressure, making it difficult to identify the
liquid-vapor interface. The resulting vapor pressure under the operating
conditions of the simulations is roughly half an atmosphere.  In a previous
paper \cite{p1} we considered tetramer molecules as well, which have
negligible vapor pressure, and showed that in
contrast to macroscopic drops \cite{nagel} the presence of vapor is irrelevant in
nanodrop surface impacts.  The results here pertain to dimers but the
general characteristics of their behavior is expected to be general.
When we discuss variations in the local energy density below, the explicit
numbers refer to the sum of the LJ and FENE energies in Eq.~\ref{eq:ljf};
we are concerned only with energy variation  within the drops so that the 
choice of the zero of energy is irrelevant.  

In the remainder of the paper lengths are given in terms of the approximate
LJ atomic diameter $\sigma$ and times in terms of the natural time unit
$\tau=\sigma(m/\epsilon)^{1/2}$ where $m$ is the common mass of all atoms.
Energies are expressed in term of $\epsilon$ and temperatures in terms of 
$\epsilon/k_B$, where $k_B$ is Boltzmann's constant.
For physical estimates of dimensionful quantities we use LJ parameters for 
argon, $\sigma\approx 0.34$nm, $\tau\approx 2$ps and $\epsilon/k_B\approx
120^\circ$.

Fluid atoms move according to Newton's equation with a force given by the 
gradient of the potentials above, integrated by Gear's method
with a time step of 0.0005$\tau$, a relatively small value to allow for
the significant collisional velocities. 
The particle motion is found by computing the net force exerted on the
particle's atoms by neighboring fluid atoms, and integrating Newton and
Euler's equations for rigid bodies, in quaternion variables \cite{at}. 

To prepare the drops, we begin with a cubic lattice of atoms, and first
select all atoms within a radius 36$\sigma$ of a center.  Within this 
spherical region, initial particle centers are located at uniformly 
distributed sub-lattice sites, and atoms with a radius 2$\sigma$ of each
center are assigned to a particle.  The resulting particles contain 32 atoms
each and have a rather ragged appearance when each atom is depicted as a
sphere, but an examination of the time-averaged liquid density field around 
a particle in equilibrium shows that the effective particle is in fact 
spherical with a radius of approximately
2$\sigma$.  For liquid marbles, the particle centers are instead 
located at uniformly spaced positions on the surface of the fluid sphere.  
The composite drop is placed at the center of a periodic box of dimensions
$(X,Y,Z)=(300,100,300)\sigma$, and initially given random 
velocities normalized to a temperature in the solid phase, $T=0.2$. 
The temperature 
is then ramped up into the liquid-vapor coexistence region at $T=0.8$ over
a time 100-300$\tau$, using a Nos\'e-Hoover thermostat. This ramp 
procedure allows the drop to melt into a well-define liquid sphere
surrounded by vapor without violent shape 
fluctuations.  Subsequently, the drop is equilibrated at $T=0.8$ until the
shape and the average particle distribution stabilize, typically requiring
500-1000$\tau$.  

\begin{figure}[h]
\begin{center}
\includegraphics[width=0.3\linewidth]{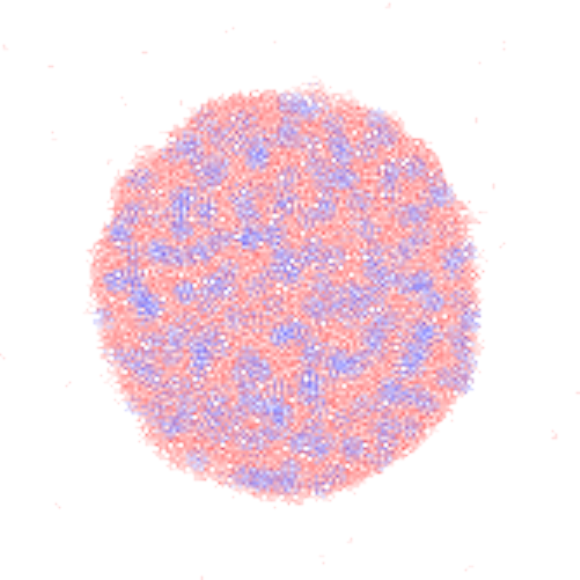}
\includegraphics[width=0.3\linewidth]{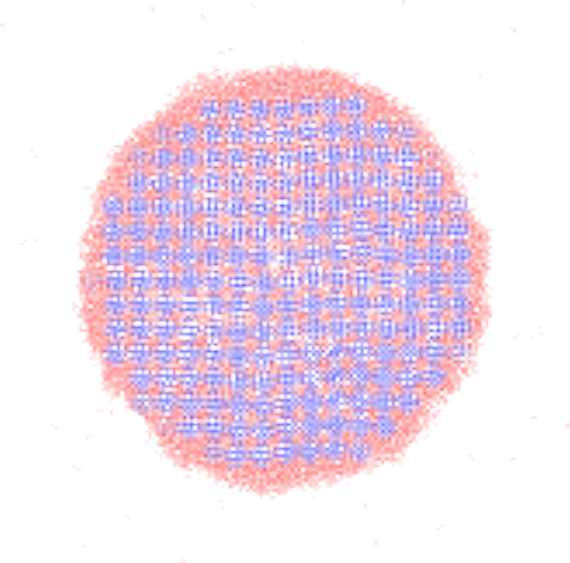}
\includegraphics[width=0.3\linewidth]{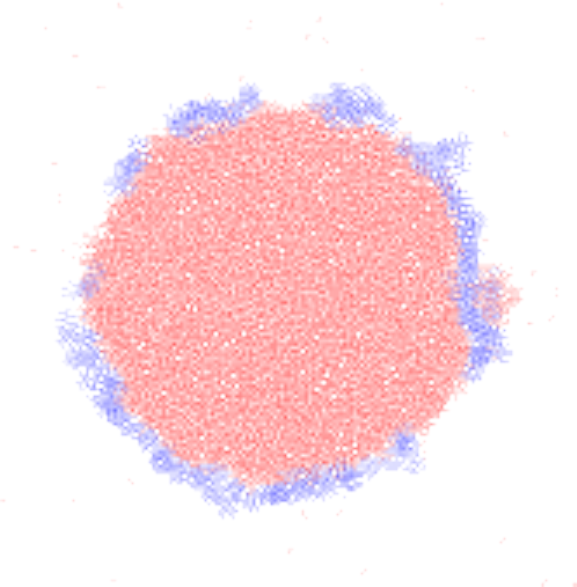}
\vspace*{0.1in}
\includegraphics[width=0.3\linewidth]{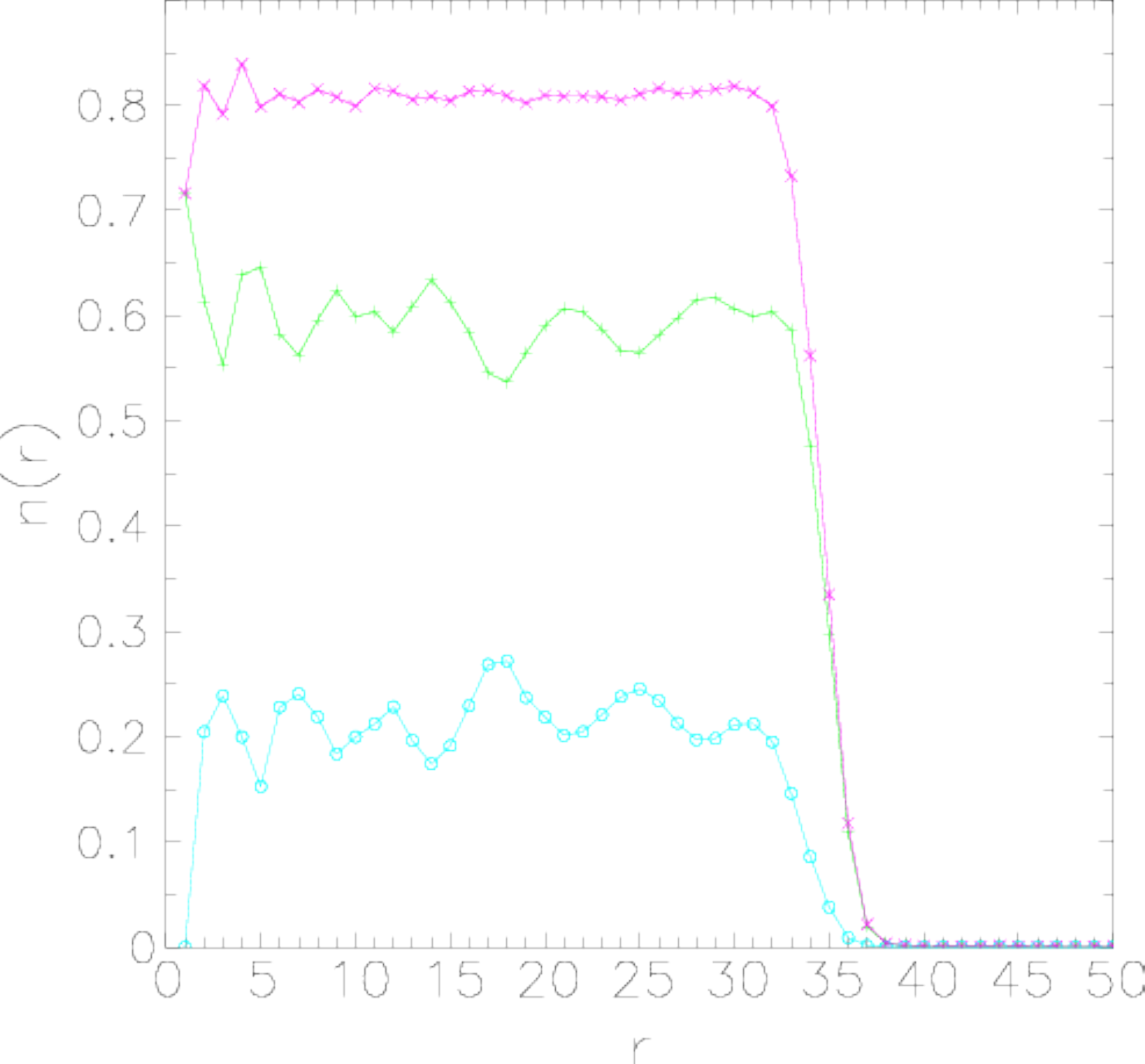}
\includegraphics[width=0.3\linewidth]{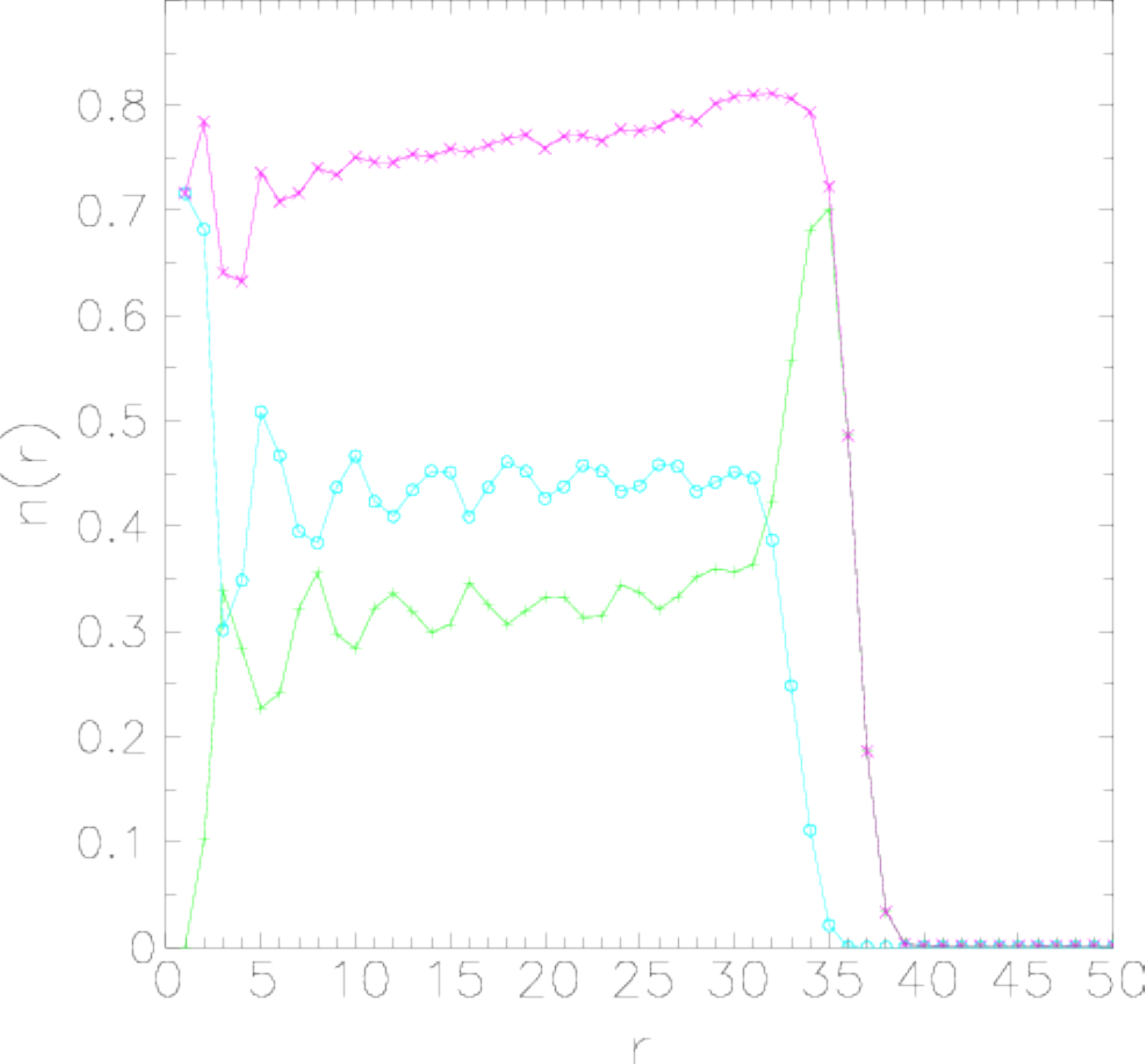}
\includegraphics[width=0.3\linewidth]{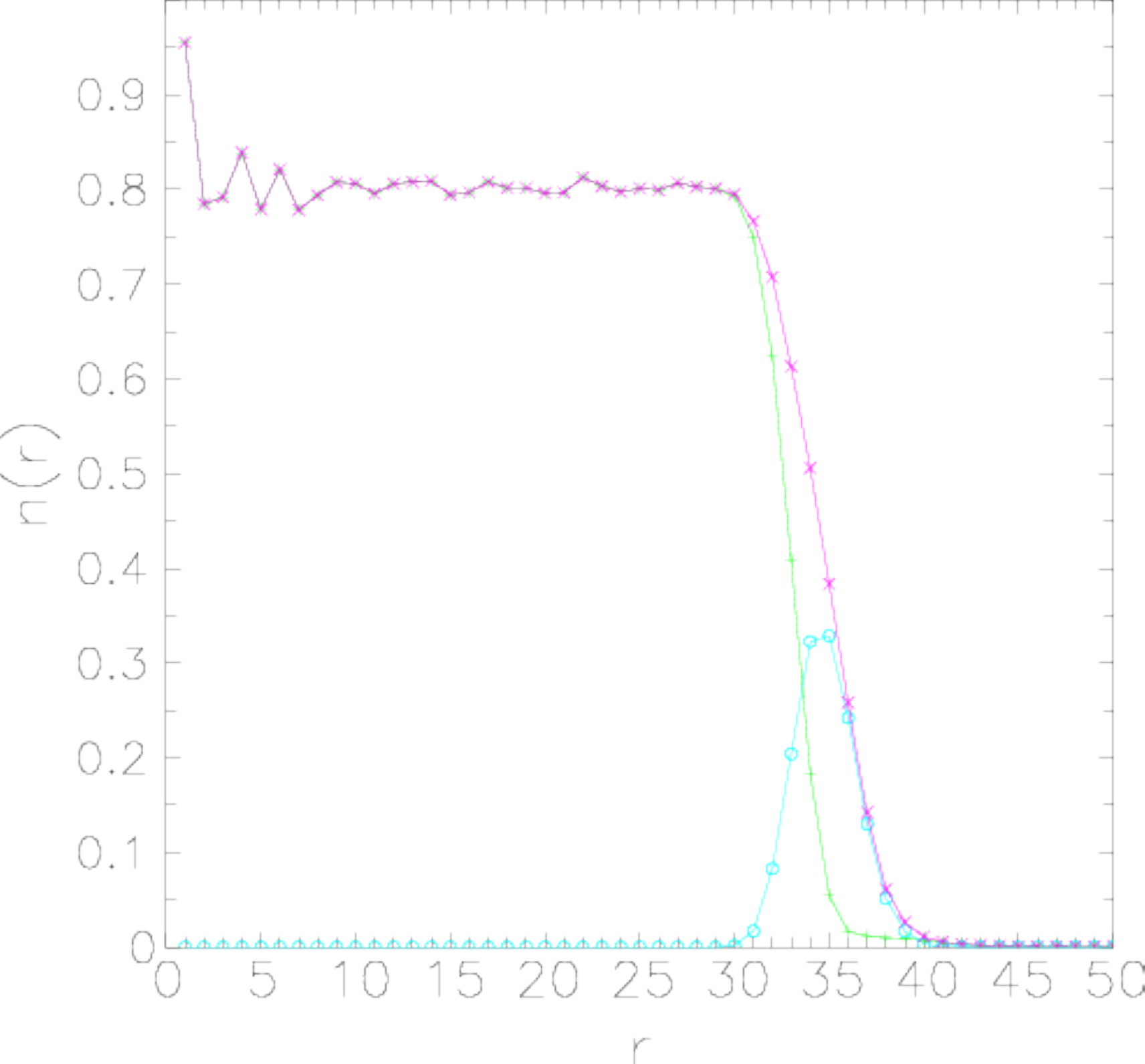}
\caption{Top row, left to right: snapshots of a $10\sigma$ 
section through the center of the 21\%, 42\% and marble drops. Fluid and 
particle atoms are shown as red (light) and blue (dark) dots.  
Bottom row, left to right: radial
density plots of the same three drops.  Particle density is shown in cyan 
(``o''), fluid density in green (``+'') and total density in magenta (``x'').}
\label{fig:drops}
\end{center}
\end{figure}

The structure of the drops is displayed in Fig.~\ref{fig:drops}, where for
each case we show a snapshot of a slab of width $10\sigma$ through the 
center along with the total, fluid and particle radial densities $n(r)$,
at the end of the equilibration period.
The particles in the two suspension drops are approximately uniformly
distributed but those in the liquid marble are not, and show some
unavoidable spatial variation along the surface.  The reason is that 
the marble particle atoms
must have a weaker attraction to the liquid to remain on the drop surface
and are consequently more mobile. The only other variable parameter is the
surface density of particles, but at lower density they are even more mobile
while at higher density they tend to leave the surface.
The number of fluid atoms and (32 atom) particles in each drop is given 
in Table I,
along with the nominal radius, defined as the point where the total density 
falls to half the value at the center.  We also indicate the number of atoms
and radius of a pure liquid dimer drop used for comparison below.
The drop diameters and number of atoms are similar in all cases, but not 
precisely the same.

\begin{table}
\begin{center}
\begin{tabular}{|c||c|c|c|c|} \hline
        	& liquid  & 21\%    & 42\%  & marble \\ \hline\hline
fluid atoms     & 157216  & 106638  & 90770 & 125470 \\ \hline
particles       & --      & 1048    & 2043  & 649    \\ \hline
radius          & 36.0    & 34.9    & 36.2  & 35.0   \\ \hline
\end{tabular}
\label{table1}
\caption{Drop parameters}
\end{center}
\end{table}

\begin{table}
\begin{center}
\begin{tabular}{|c||c|c|} \hline
$u_0$   & $Re$  & $W\!e$\\ \hline\hline
1.0     & 10.3  & 56.5  \\ \hline
2.0     & 41.2  & 226.  \\ \hline
3.0     & 31.0  & 508.  \\ \hline
\end{tabular}
\label{table2}
\caption{Flow parameters.}
\end{center}
\end{table}

For surface impact simulations, a denser slab of {\it fcc} solid, with lateral
dimensions $(X,Z)=(300,300)\sigma$ and one unit cell thickness in $Y$,  
containing 147,456 atoms, is 
placed at the base of the simulation box. These solid atoms are tethered to
their lattice sites using a spring of stiffness 100$\epsilon/\sigma$,
allowing small thermal fluctuations in position. The solid is equilibrated at 
the same temperature $T=0.8$ as the equilibrated fluid.  Impact is initiated 
by turning off the 
fluid thermostat and giving each atom in and near the drop a fixed downward
velocity.  The solid temperature is fixed throughout the simulation by the 
thermostat, corresponding physically to impact on a material of very high 
thermal conductivity held at fixed temperature.  For drop collision 
simulations, after the
fluid has equilibrated, the box is doubled in size and a duplicate atomic 
system with the same velocities and {\em relative} positions is placed in 
the open region just above the original drop.  A further (short)
equilibration interval allows the vapor in the region between the drops 
to mix uniformly, and then the two drops are given equal and opposite
velocities so that they collide.  The latter instant is defined as $t=0$
in the subsequent results.

The Reynolds and Weber numbers for the {\em liquid} drop at the collision
velocities $u_0$ studied in this paper, $Re=\rho u_0 R/\mu$ and 
$W\!e=\rho u_0^2 R/\gamma$, are given in Table II.  The viscosity 
$\mu=2.80 m/\sigma\tau$ and the surface tension $\gamma=0.51m/\tau^2$, of
the pure fluid are 
determined in separate simulations of Couette flow
and a planar liquid vapor interface for the same liquid drop, as discussed 
in \cite{p1}. The latter paper calculates the sound speed for the 
dimer liquid studied here, $c_s\approx 4\sigma/\tau$, and the resulting Mach
number $Ma=u_0/c_s$ ranges from 0.25 to 0.8.  The presence of particles 
of course alters the viscosity and surface
tension of the drops, so these numbers should be thought of as convenient 
non-dimensionalization of the velocity but not exactly the 
ratios of accurate physical length or time scales.

In the remainder of the paper we display the results of only a single simulation
for each value of particle concentration and impact velocity, but in fact a
number of additional cases with small differences in drop size, velocity
and interactions were carried out as well.  The differences preclude simple
averaging of the results, and in any event there are too few runs available
to constitute a reliable statistical sample, but the qualitative results
described below here are robust in the sense that the behavior is not
significantly altered if a different realization of the initial conditions 
or a small variation in parameters is considered.

\section{Surface impact} \label{impact}

\begin{figure}[h]
\begin{center}
\includegraphics[width=0.9\linewidth]{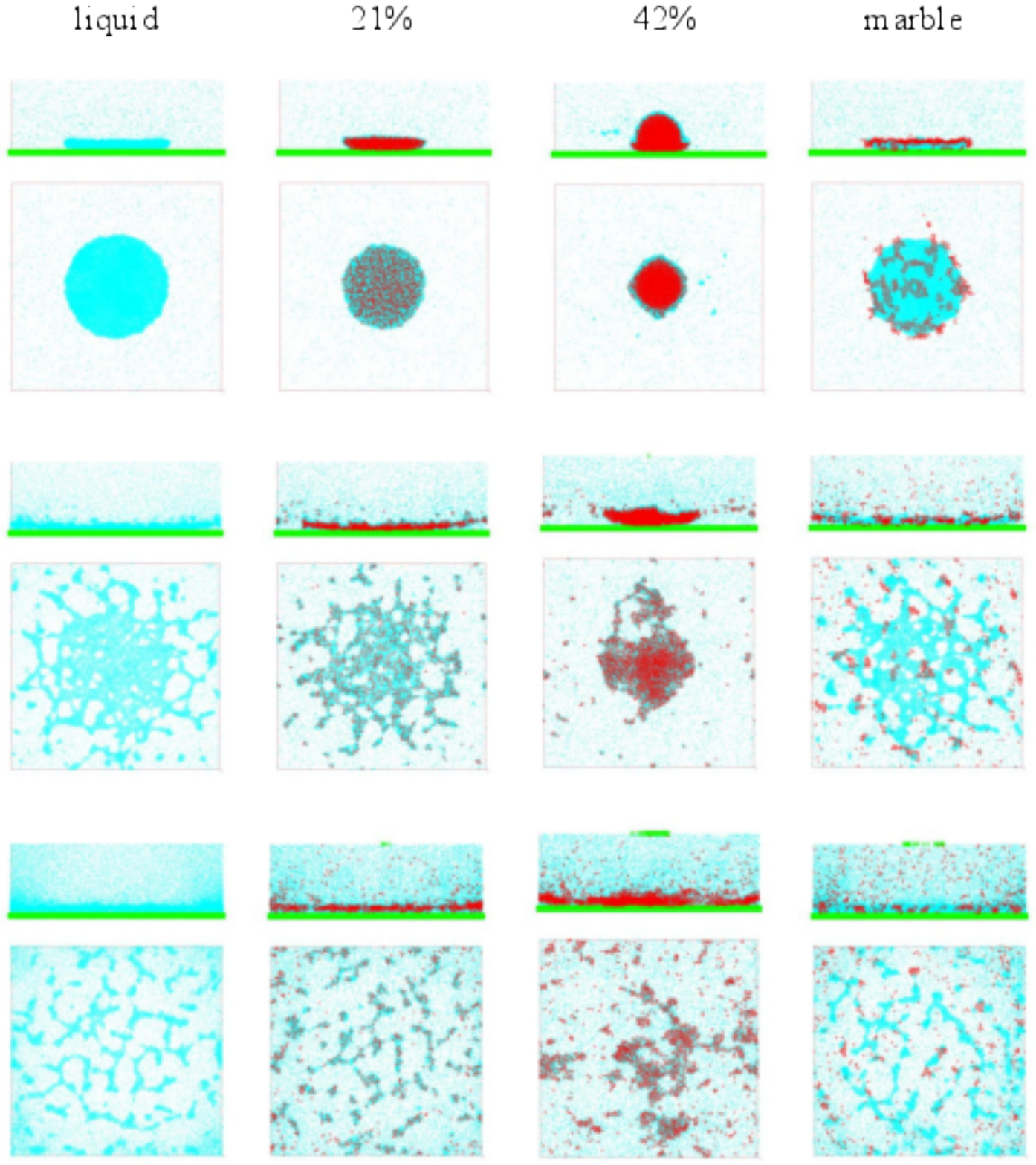}
\caption{Snapshots of different drops impacting a non-wetting
surface at time 100$\tau$.  Columns, left to right: pure liquid, 21\%
suspension, 42\% suspension and liquid marble.  Rows, top to bottom: initial
velocities $u_0=1, 2, 3\sigma/\tau$. Fluid and solid atoms shown in cyan and
red, respectively.}
\label{fig:nw_impact}
\end{center}
\end{figure}

\begin{figure}[h]
\begin{center}
\includegraphics[width=0.9\linewidth]{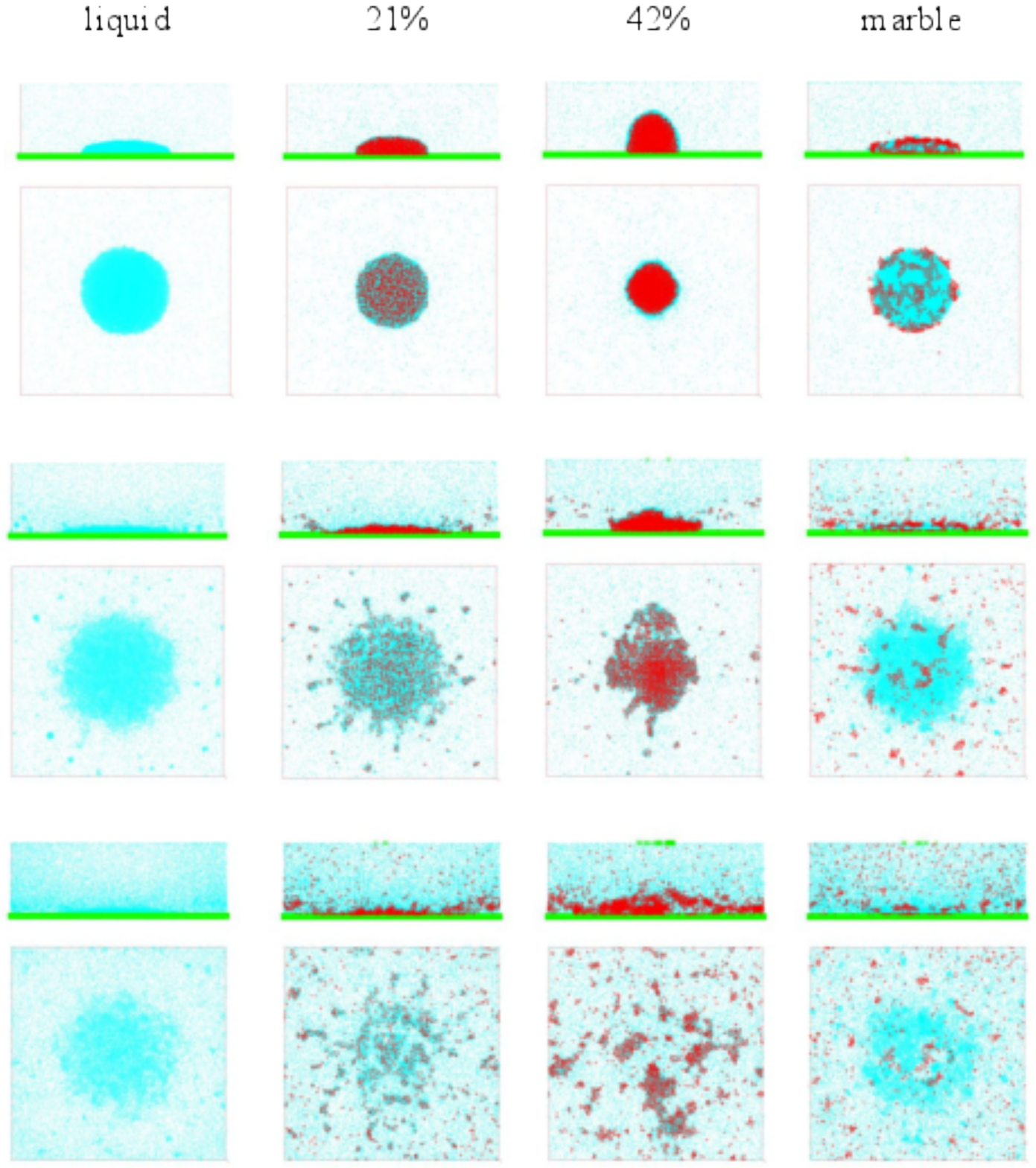}
\caption{Side view plus top view snapshots of different drops 
impacting a wetting
surface at time 100$\tau$.  Columns, left to right: pure liquid, 21\%
suspension, 42\% suspension and liquid marble.  Rows, top to bottom: initial
velocities $u_0=1, 2, 3\sigma/\tau$. Fluid and solid atoms shown in cyan and
red, respectively.}
\label{fig:wet_impact}
\end{center}
\end{figure}

In this  section we discuss the impact of particle-laden drops on atomistic
solid surfaces which are homogeneous, planar up to thermal fluctuations and  
are either completely wetting or completely non-wetting to the liquid and
solid atoms within the drop.  For reference we recall the behavior of pure
liquid drops \cite{p1}; a snapshot of the two wettability cases at time
100$\tau$ is shown in the left-hand columns of Fig.~\ref{fig:nw_impact} and
Fig.~\ref{fig:wet_impact}, respectively, for different initial velocities.  
At $u_0=1.0\sigma/\tau$ the drop deforms and 
spreads laterally after impact without emitting significant vapor.
At later times, a drop on a non-wetting surface contracts, lifts off the
surface and eventually regains its spherical shape, whereas on a wetting
surface the drop after its initial deformation continues to spread but at a 
much slower rate.
At $u_0=2$ the drop initially forms a crown splash and subsequently
disintegrates, producing extensive vapor and leaving a ramified pattern
resembling spinodal decomposition in the non-wetting case but a fairly
well-defined circular puddle on a wetting surface.  At $u_0=3$ the drop
disintegrates immediately and fills the simulation box with vapor, while
the surface is left with tendrils of liquid in the non-wetting case or an
irregular puddle on a wetting surface.  At times later than shown in the
latter two cases, the vapor becomes uniformly distributed in space (the 
simulation domain is a periodic box) while the
liquid remaining on the surface aggregates into larger droplets.
 
\begin{figure}[h]
\begin{center}
\includegraphics[width=0.9\linewidth]{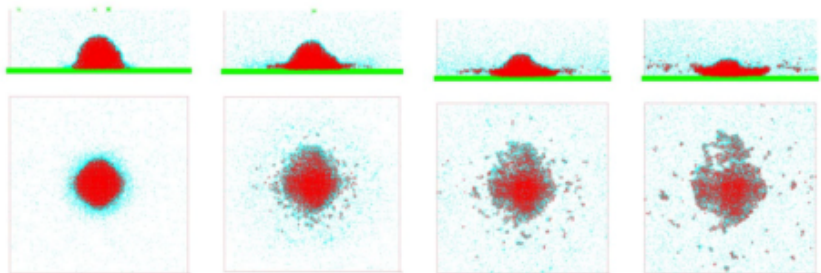}
\caption{Impact of a 42\% suspension drop on a non-wetting surface at
velocity $u_0=2\sigma/\tau$, at times
(left to right) 20, 35, 50 and 75$\tau$.  For 100$\tau$ see
Fig.~\ref{fig:nw_impact}.} 
\label{fig:seq_42}
\end{center}
\end{figure}

An example of the time evolution of the impact of a particle-laden drop is 
given in Fig.~\ref{fig:seq_42}, for the case of the 42\% suspension falling on a
non-wetting surface at $u_0=2\sigma/\tau$.  (Analogous plots for pure
liquid drop impact are given in \cite{p1}.)  After contacting the solid
surface a fluid rim forms
(20$\tau$), and then a lamella of fluid and isolated single particles 
proceeds radially outward along the surface, accompanied by a 
crown splash (35$\tau$).  The lamella breaks into fragments and particles 
begin to cluster (50$\tau$) and then the body of the drop breaks up
(75$\tau$).  At 100$\tau$ the fragments begin to coarsen, and subsequently
the solid structure becomes smoother while the fragments move off.  While
the time evolution of the drop differs in detail from case to case, we note
that the
fate of the drop is evident at time 100$\tau$, so we summarize the results 
of the various impact simulations by displaying their state at this time, in
Figs.~\ref{fig:nw_impact} and \ref{fig:wet_impact} for now-wetting and wetting
surfaces, respectively.

The behavior of the 21\% concentration suspension is rather similar to that
of pure liquid, the principal differences being a reduction in the degree of
lateral spreading, and a more attenuated surface residue of liquid along with
more vapor in the higher-velocity impacts.  At later times, the low velocity
drop recoils from the surface and becomes spherical while the other cases
continue to produce vapor while their surface residue coarsen. The term 
``coarsening'' here is used in further analogy to spinodal decomposition 
and indicates that the thicker liquid tendrils on the surface accumulate
molecules and particles and thicken while the thinner ones evaporate.  
The reduction in the spreading diameter may be attributed to the increased 
viscosity of a suspension in comparison to pure liquid. The additional
vapor and stringier surface residue at the higher impact velocities, in
comparison to pure liquid, we
attribute to the clustering of liquid around the particles,
whose atomic interactions with the liquid have the same strength as the 
liquid-liquid ones in this case.

The higher concentration (42\%) suspension shows distinctly different behavior.
At low velocities a nearly rigid bowler hat shape is formed, in which the 
solid particles first striking the surface are directed outwards and begin 
to spread laterally but the bulk of the drop acts as a semi-rigid body and
retains its original shape. In fact the lateral motion ceases at around
time 70$\tau$, while the drop deforms the solid surface downward, and 
on recoil the drop leaves the surface with little further change in shape.
Similar behavior is observed in experiments involving the impact of drops of
yield-stress fluids \cite{german,luu}.
At a higher velocity, $u_0=2.0\sigma/\tau$, the drop is strongly deformed into
an irregular and partially fragmented saucer shape, with less vapor and more
liquid retained in the saucer.  The particles in a sense act as a glue 
holding the drop together.  The drop breaks up completely only at the highest
velocity, leaving a surface residue in the form of small suspension
clusters. At later times the higher velocity impact states coarsen in part: 
smaller fragments of solid become smoother while remaining in place or else 
merge with the larger fragments, while the larger ones grow in volume while 
becoming more compact as they accumulate liquid atoms and smaller solid 
clusters.

We have also considered the situation where there is {\em no} attractive
potential between the solid atoms in different particles and only a 
short-distance
repulsion. At low impact velocity $u_0=1.0\sigma/\tau$ the particles move
with the fluid, and the result resembles the 21\% case, but at higher
velocities the particles shoot out horizontally and accumulate at the edges 
of the simulation box, so that the drop's disintegration is more violent 
than at lower particle concentration.  A recent experiment observed the
ejection of a monolayer of particles in the surface impact of a suspension
drop \cite{jaeger2}, but in these simulations several particle layers are seen.
The contrasting behavior may originate from different interactions as well
the size difference, but we do not pursue the matter here.   

The liquid marble's behavior is roughly intermediate between fluid and 21\%
suspension, both in spreading diameter at low velocity and in the amount of 
vapor produced in faster impacts.  The impact tends to dislodge the
particles from the drop surface leaving them isolated, because
in this case the atomic interactions particle and liquid atoms are weaker 
than those of the liquid. On the whole, the particles make little difference 
to the behavior of the drop in impact, and the slightly smaller amount of
vapor and surface tendrils generated is the results of having fewer liquid
atoms in the original drop.

The 100$\tau$ state of the same set of drop impacts on a {\em wetting} surface 
is shown in
Fig.~\ref{fig:wet_impact}.  At the lowest velocity there is an obvious
difference in the shape of the rim at the surface:  the drop atoms are now
pulled towards the surface to produce a downward rather than upward-facing 
curvature.  Otherwise, all four drop shapes and lateral extent are similar 
to those on a non-wetting surface.  (See \cite{p2} for a discussion of low
velocity impacts on intermediate wettability surfaces.) When these cases 
are continued to longer 
times, the liquid spreads slowly while the particle-laden drops show 
negligible change in shape.  A second qualitative difference with respect to
a non-wetting surface is that in all of the higher-velocity impacts a 
residual circular saucer-shaped liquid puddle remains on on the surface.  
Both differences simply reflect
the attraction of the surface atoms, absent in the non-wetting case.
Furthermore, this attraction suppresses the formation and coarsening of the 
spinodal-like liquid tendrils seen in that situation, by acting to hold the 
liquid in place.
For either type of surface, at higher impact velocity the particles form
smaller droplets mixed with liquid at low concentration, tend to disperse 
in the marble case because of their weaker attraction to the liquid, and form 
nearly rigid irregular clusters at high concentration.  
At times beyond that shown in the figure, the puddles persist and become
somewhat more regular, the vapor tends to becoming uniform in space, and 
in the concentrated suspension the solid clusters tighten up and exhibit 
some smoothing at the extremities.

\section{Drop collisions} \label{collision}

A head-on collision of two drops has a superficial similarity to a normal 
collision on a solid surface, in the (continuum) sense that one might
imagine that the colliding
drops should be mirror images of each other with the mid-plane playing a
role similar to that of a reflecting boundary. In reality, the atoms in a
solid surface respond to those of an incoming liquid drop in a rather
different manner than those in a second drop, and it is of interest to
compare the two processes.  We first consider the collisions of two drops
which are ``identical'' in an average sense, which is to say they have the
same number of atoms and were prepared (equilibrated) in the same way.
The  two drops would then have the same density, mean radius, interfacial
thickness, etc., but distinctly different atomic positions and momenta.
In practice we cheat slightly: we duplicate a single equilibrated drop
and translate it to a position just above the original, and then shift the
atomic velocities in the two drops by equal and opposite amounts so as to
make them collide.  Note that the top and bottom of the original drop have
relative positions and velocities which differ in detail (although they have
the same average) and it is these which control the initial collision
process.  If instead we had {\em reflected} the original drop configuration 
about a mid-plane before shifting velocities the collision would have involved 
two unrealistic mirror image drops.
 
\begin{figure}[h]
\begin{center}
\includegraphics[width=0.9\linewidth]{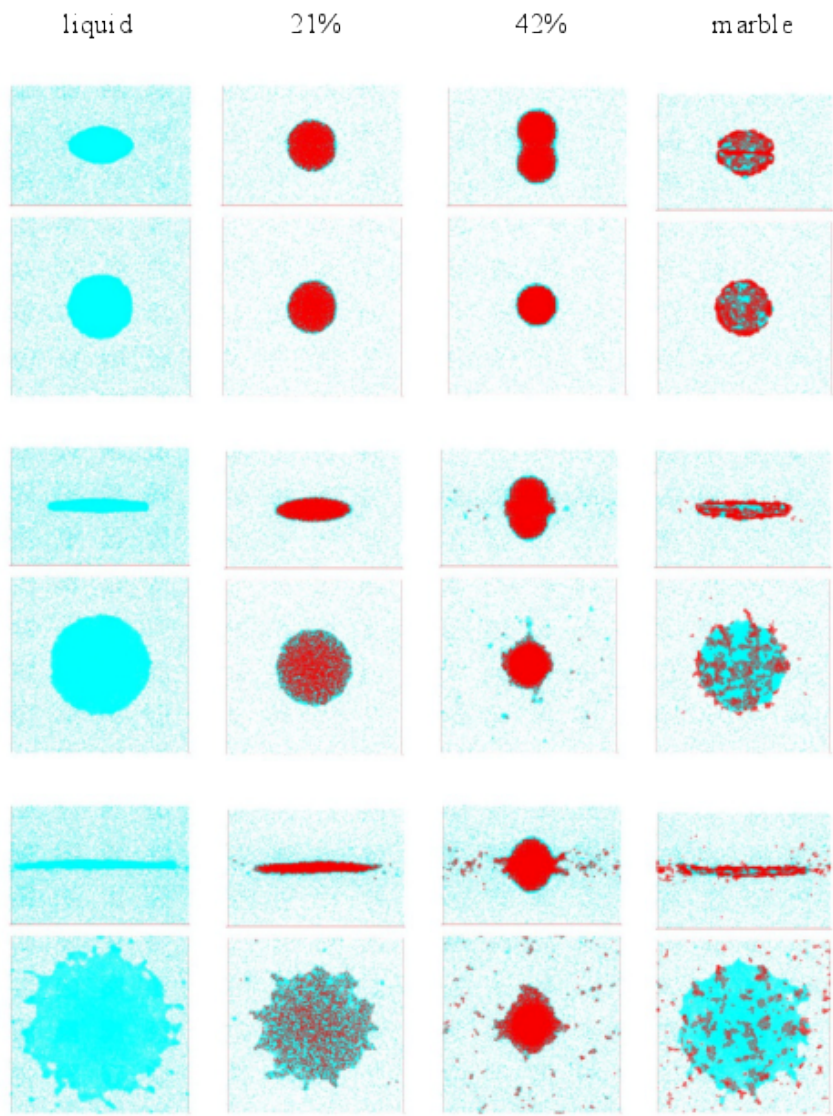}
\caption{Snapshots of head-on collisions of drops of the same 
material at time 100$\tau$.  Columns, left to right: pure liquid, 21\%
suspension, 42\% suspension and liquid marble.  Rows, top to bottom: initial
relative velocities $u_0=1, 2, 3\sigma/\tau$. Fluid and solid atoms shown in 
cyan and red, respectively.}
\label{fig:same_coll}
\end{center}
\end{figure}

The results of collision simulations of the four types of drop whose surface
impacts were considered in the previous section is shown in
Fig.~\ref{fig:same_coll}, at the same three {\em relative} velocities.
When two identical pure liquid drops collide (first column) their respective  
atoms attract each other and the drops merge.  At relative velocity 1.0 and
2.0$\sigma/\tau$ the drops merge smoothly:  once the drop's atoms are within
interaction range a neck forms and thickens, the drops continue to move
towards each other and combine into a peanut shape which subsequently evolves 
into a prolate ellipsoid, which in turn eventually relaxes to a
(larger) sphere under the action of surface tension.  In this case, and also
if the solid
concentration is not too high, there is a decaying oscillation about the
final spherical shape. The effect of
increasing the initial relative velocity is to increase the initial
compression giving a disk-like intermediate state.  When the initial
velocity is further increased to 3$\sigma/\tau$ the disk is thinner and
becomes rather ragged at its edges;  in this simulation it reaches the edge
of the simulation box and effectively merges with its periodic neighbors,
but in open space it would emit small secondary droplets from the rim while
the main body eventually contracted to a sphere.  At still higher collision
velocities (not shown) a merged disk forms but spreads so rapidly as to open
holes and then disintegrates into smaller drops. These collisions  may be 
compared
to impact on a wetting (attractive) surface: there is a visual similarity 
between half of the lower-velocity collision shapes here and the impacting 
drop at velocity 1.0$\sigma/\tau$, but the variation of drop radius and height 
with time is not the same.  More significantly, an impact at velocity 2.0
causes a drop to disintegrate whereas in collisions the drop retains its
integrity even at relative velocity 3.0.  Physically, atoms in a solid are 
constrained to have small displacements and tend to cause impinging liquid 
molecules to reflect and transfer their kinetic energy back to the incident 
drop, whereas atoms in a second drop are not so constrained and are capable
of absorbing and redistributing the incident energy.  A more detailed
analysis of energy transfer is given in the next section.

\begin{figure}[h]
\begin{center}
\includegraphics[width=0.9\linewidth]{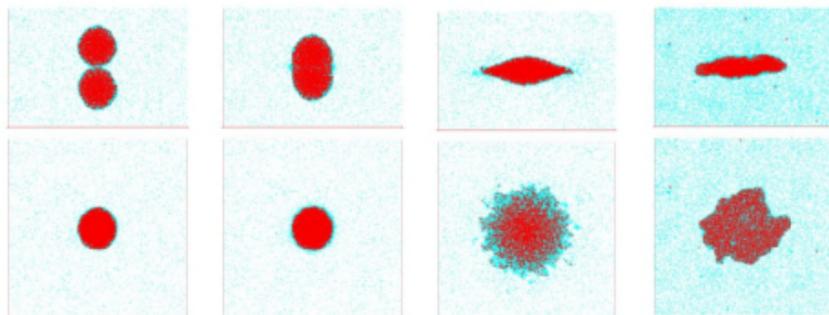}
\caption{Collision of two identical drops of a 21\% suspension at relative 
velocity 3$\sigma/\tau$ at times (left
to right) 10, 20, 50 and 500$\tau$; the 100$\tau$ snapshot is included in 
Fig.~\ref{fig:same_coll}.}
\label{fig:seq_21}
\end{center}
\end{figure}

The same collision protocol for the various particle laden drops gives the 
states at time 100$\tau$ shown in the remaining columns of 
Fig:~\ref{fig:same_coll}.  The 21\% suspension drops merge smoothly at the
lowest collisional velocity but at a
much slower rate, going
from touch to peanut to sphere (in the figure) to slightly prolate ellipsoid 
(later), and continues to contract at 1000$\tau$.
Relaxation to the final spherical shape requires much longer
times.  At velocity 2$\sigma/\tau$, the behavior is similar but the time
scale differs: the drops merge and contract into an ellipsoid, and then
begin to relax back after 150$\tau$, but remains highly ellipsoidal even at
1000$\tau$.  At both of these impact velocities, the particles 
remain uniformly
distributed in the drops' interiors.  At initial velocity 3$\sigma/\tau$,
however, a thin disk of merged liquid forms having an irregular rim and with
some particle and secondary droplet emission.  Subsequently, the edges
smooth due to rearrangement and the drop begins to relax to a sphere, but
very slowly and again remains highly disk-like at 500$\tau$.  
Snapshots of the detailed time evolution for this case are shown in
Fig.~\ref{fig:seq_21}.  At still higher
collision velocity a thin disk with irregular edges forms, which again develops
holes as it expands to each the box boundary and emits small particle
clusters.

The 42\% suspension again shows a semi-solid behavior.  At incident velocity
1.0$\sigma/\tau$ a partial merger occurs in the contact region, leading to a
peanut shape after about 70$\tau$, which shows almost no subsequent shape 
variation aside from a very weak contraction. 
More precisely, if we examine the location of
the center of mass of the original liquid and solid atoms in the two drops,
after colliding the average position of the
solid atoms hardly changes, whereas that of the liquid atoms contracts 
very slowly ($\sim \pm 0.05\sigma/\tau$) as they rearrange inside the 
solid structure.  At velocity 2$\sigma/\tau$ the original drops
move closer together after contact than in the previous case, leading to a
rough oblate ellipsoid, but again 
nearly freeze in place after 70$\tau$.  Here the collision produces an
outgoing liquid sheet in the collision plane, which breaks up into droplets.
A small rim of particles is carried along and several additional small
particle clusters are emitted.  At the higher velocity 3$\sigma/\tau$, the
same behavior is more evident, with a stronger emission of particle and
fluid along the mid-plane, leaving an irregularly-shaped nearly-rigid
suspension drop behind. A simulation at still higher velocity 4$\sigma/\tau$ 
gives a qualitatively similar result, with more emission and irregularity. 

The liquid marble collisions at low velocity 1.0$\sigma/\tau$ allow the
particle to remain mostly at the interface, where they have the effect of
squeezing the liquid into a thinner and broader disk than a pure liquid 
drop, presumably due to their inertia, although some particles are knocked
loose by the collision.  At higher velocities many particles are dislodged
from the surface of the drop, and their outward motion broadens and distorts
the edges of the liquid-filled disk at $u_0=2\sigma/\tau$ and tears it apart 
completely at velocity $3\sigma/\tau$.  A recent experiment involved the
collision between a moving, densely-covered liquid marble and a second, much
larger one at rest \cite{planchette}.  At low velocities the moving marble 
separates after impact while at higher velocites it is absorbed.  However,
the system studied is somewhat different from the present simulations;  aside 
from the degree of surface coverage, the experimental marbles are large 
enough for gravity to play a role in their behavior.

\begin{figure}[h]
\begin{center}
\includegraphics[width=0.9\linewidth]{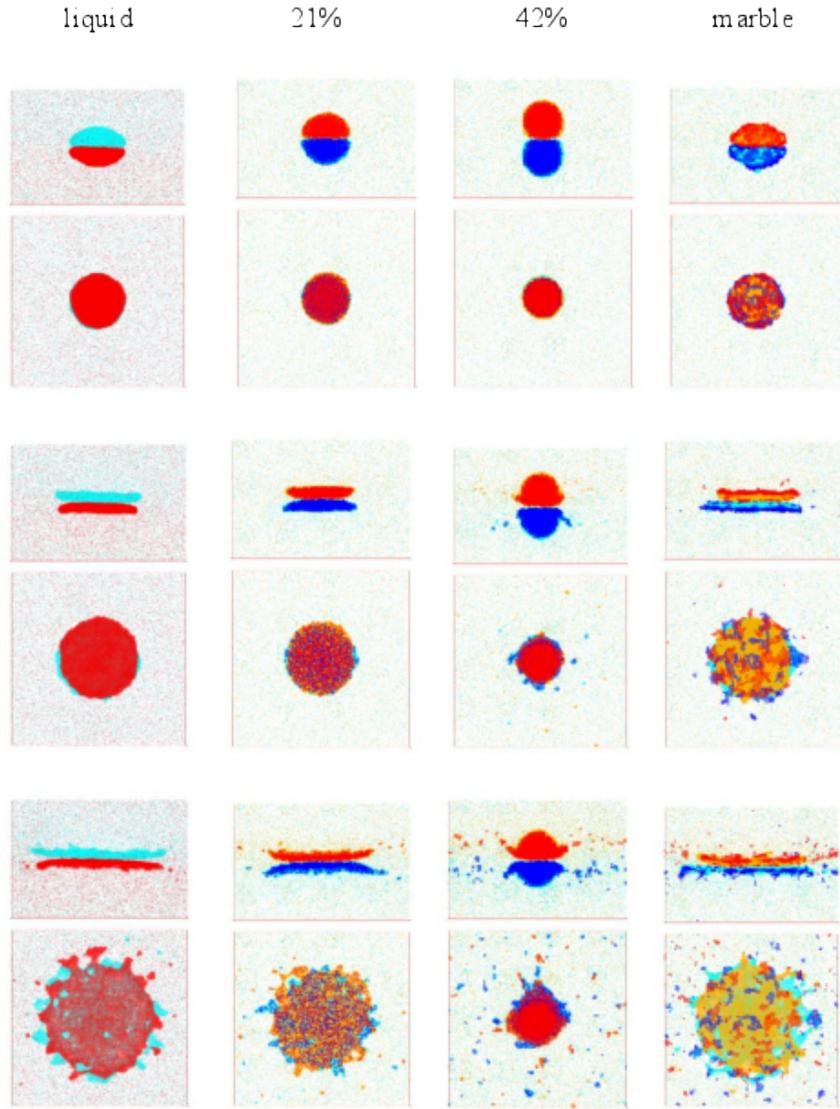}
\caption{Snapshots of head-on collisions of drops of the
``opposite'' material, as define in the text,  at time 100$\tau$.  
Columns, left to right: pure liquid, 21\%
suspension, 42\% suspension and liquid marble.  Rows, top to bottom: initial
relative velocities $u_0=1, 2, 3\sigma/\tau$. Fluid and solid atoms shown in 
cyan and red, respectively.}
\label{fig:opp_coll}
\end{center}
\end{figure}

\begin{figure}[h]
\begin{center}
\includegraphics[width=0.9\linewidth]{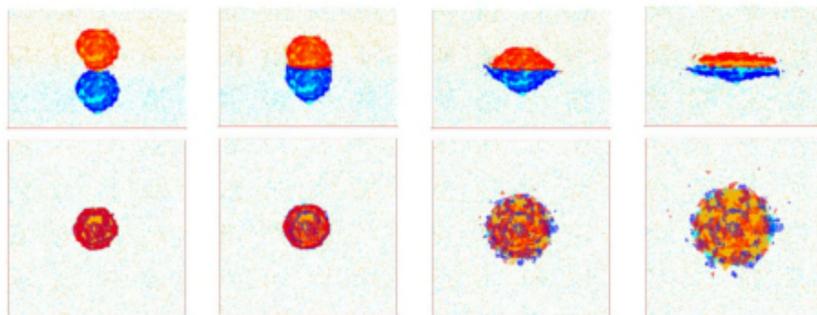}
\caption{Collision of two ''opposite'' drops of a liquid marble at relative
velocity 2$\sigma/\tau$ for times (left
to right) 15, 30, 50 and 75$\tau$; 100$\tau$ is included in 
Fig.~\ref{fig:opp_coll}.}
\label{fig:seq_marb}
\end{center}
\end{figure}

Just as collisions of drops of the same material has a superficial
similarity to impact on a wetting surface, we can compare impacts on
non-wetting surfaces to collisions between drops of ``opposite'' material by
choosing the atoms of the respective drops to have only a repulsive
interaction.  In practice we set the coefficient $c$ in Eq.~\ref{eq:ljf} to
zero for interactions involving both fluid and particle atoms in the 
different drops, and repeat the same suite of 
collision simulations as above.  The
results, again at time 100$\tau$, are given in Fig~\ref{fig:opp_coll}.    
The obvious qualitative difference is that the drops stand each other off
rather than merge, and even at the highest velocity when the drops
disintegrate, the fragments mostly occupy their original half-spaces.
A concrete illustration of the collision process for this case is given in
Fig.~\ref{fig:seq_marb} for the case of two ``opposite'' marbles at relative
velocity 2$\sigma/\tau$.  Note that
the particles remain attached to the drops until they become severely
deformed, and even then only a few are expelled, a quite different behavior 
from that of a surface impact at this velocity.
In most cases the structure after collision has the same general stability 
as collisions between drops of the same material 
in terms of the smoothness of the drop/vapor boundary and the degree to which
particles are expelled.  The principal differences are seen in the mid-plane 
contact region, which we discuss further in the next section.

\section{Energy transfer} \label{transfer}

\begin{figure}
\begin{center}
\includegraphics[scale=0.4]{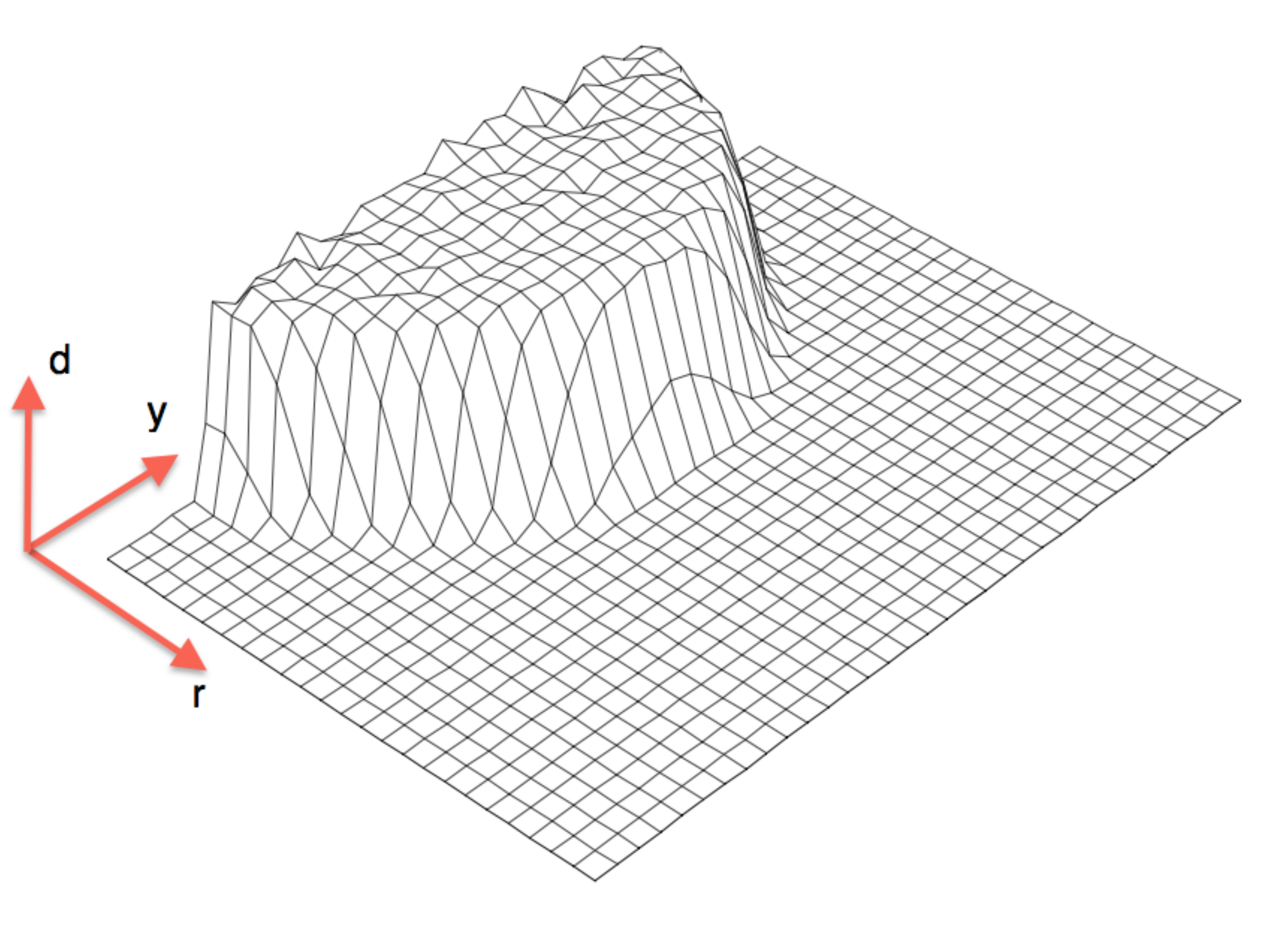}
\caption{Surface elevation plot of the two-dimensional density field $d(r,y)$ 
of an equilibrated 21\% suspension drop, in cylindrical coordinates averaged 
over azimuthal angle, where $y$ is the height above the surface and $r$ is
distance from the $y$-axis. }
\label{fig:coord}
\end{center}
\end{figure}

The translational kinetic energy of a drop is redistributed between 
the drop and the surface on impact or between the two drops in a collision, 
but the details of the division between kinetic and potential energy within 
the drop(s) and energy transferred to the solid is not obvious.  Since the
simulations track the atomic positions and momenta, the relevant information
is available, and in this section we study the local distribution of energy.
Since both normal surface impacts and head-on collisions of spherical drops
have an ostensible rotational symmetry about the collision ($y$) axis,
we adopt a cylindrical coordinate system $(r,\phi,y)$, average all fields
over the azimuthal angle $\phi$ and consider the energy fields as a 
function of ``height'' $y$ and two-dimensional radius $r$ at different times,
separately for fluid and particle species.  In Fig.~\ref{fig:coord} we
illustrate the coordinate system and the surface elevation format for the 
subsequent field 
plots, using the density field ($d$) for a pure liquid drop in equilibrium as 
an illustration. Aside from the energy, since we remarked above that the 
Mach number in these simulations is not small and fluid compressibility may 
play a role, the spatial variation of density on impact is likewise of 
interest.  Furthermore, the individual species 
densities provides information on the coherence of fluid and
particles and the degree of mixing of the species in the collision case, so 
in the next section
we present the time evolution of the individual species density fields as 
well. In fact the axial symmetry is approximate at best, and absent in all 
but a statistical sense in the higher-velocity impacts and collisions, but
the fluctuation information lost in this way is of lesser interest.
In this section, we give several examples of the evolution of various 
internal flow fields within the drop and then draw some general conclusions. 

\begin{figure}
\begin{center}
\includegraphics[width=0.9\linewidth]{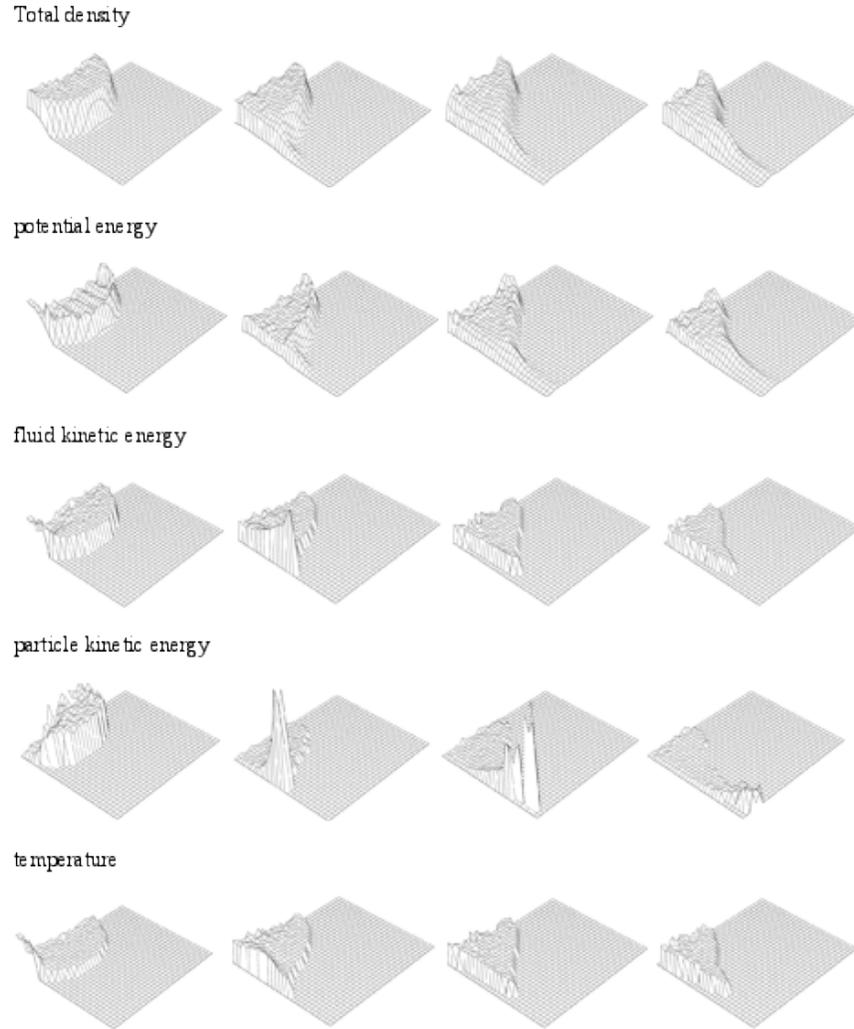}
\caption{Surface elevation plot of the two-dimensional density and energy 
fields arising
in the surface impact of a 42\% suspension drop on a non-wetting surface.
The format is that of Fig.~\ref{fig:coord} and the corresponding snapshots 
of the impact are given in Fig.~\ref{fig:seq_42}
Left to right:  times 10, 20, 30 and 50$\tau$.  Top to bottom: total density,
potential energy density, fluid kinetic energy, particle kinetic energy and
temperature.  The maximum heights in the left-hand frames are
density: 0.946$\sigma^{-3}$, potential energy: 5.66$\epsilon$, fluid kinetic
energy: 5.49$\epsilon$, particle kinetic energy: 75.35$\epsilon$ and
temperature 3.02$\epsilon/k_B$.}
\label{fig:energy_42}
\end{center}
\end{figure}

We begin with a representative surface impact case, a 42\% suspension 
drop incident on a non-wetting surface at velocity $u_0=2\sigma/\tau$.
in Fig.~\ref{fig:energy_42} we present the cylindrically-averaged total
density, potential energy, kinetic energy and temperature fields, during the
earlier stages of the impact.  Corresponding snapshots are given in 
Fig.~\ref{fig:seq_42}.  The density field exhibits fluctuations and a
slight rise at the top of the drop where fluid accumulates at first and
is later expelled into vapor, but no systematic variation in the interior 
and, in particular, only a slight 20\% transient enhancement at the bottom 
where it hits the solid surface.
The potential energy roughly follows the density and, somewhat surprisingly
for a hard collision, shows only a similar rise at the
solid surface just after contact. The subsequent decay is due to the
dispersal of fluid into vapor, leaving the particles behind in the irregular
saucer seen in the snapshot.  The ridge in the potential energy field at 
the edge of the drop appears because the atoms at the edge have fewer
interacting neighbors and carry less (negative) LJ interaction energy,
leading to an increase in the value of the energy density there.  

The effects of the impact are most apparent in the kinetic energy. The fluid
kinetic energy carried by the drop initially builds up at the impacting 
surface and is then carried outward along it by the spreading fluid 
lamella before
dispersing, partly to the remainder of the drop and partly to the solid
surface where it is conducted away ({\em i.e.}, removed by the wall
thermostat).  The particle kinetic energy is small near the surface 
immediately after impact but subsequently a packet of energetic particles move
outward along the surface as they are displaced to form the solid
saucer seen in the snapshots, and eventually dissipates as the particles
come to rest.  The local fluid temperature (the particles are athermal) 
also rises near the surface, beginning at impact and persisting throughout 
the time interval shown in the figure. The persistent enhancement in
temperature near the surface is the origin of the disintegration of liquid
drops in impact at this velocity \cite{p1}.  Note that although the
translational kinetic energy is dissipated into the vapor and the
thermostat, the temperature is the {\em fluctuation} in kinetic 
energy about the local mean, which decays due to the diffusion of momentum
on a longer time scale.  

\begin{figure}
\begin{center}
\includegraphics[width=0.9\linewidth]{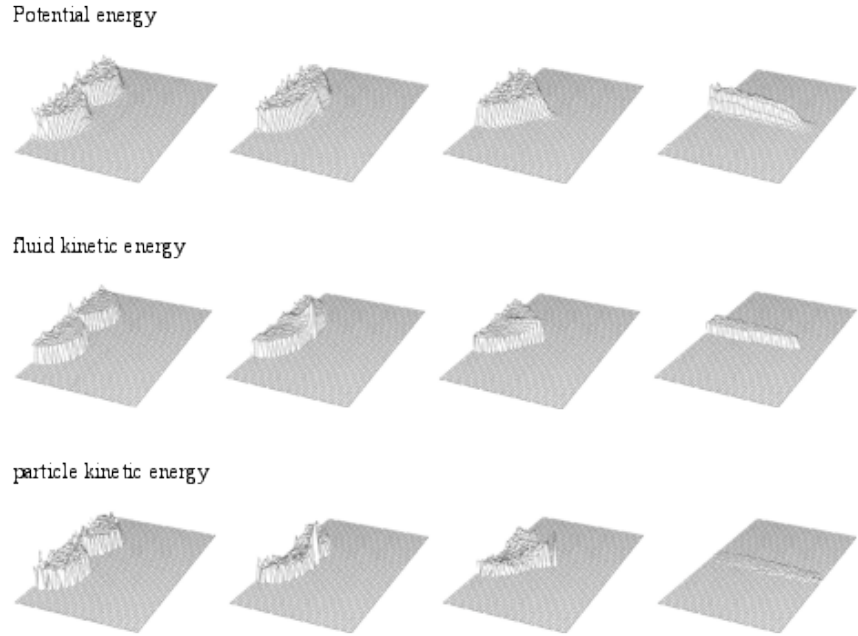}
\caption{Surface elevation plot of the two-dimensional density
fields arising in the collision of two 21\% suspension drops at velocity
$3.0\sigma/\tau$
The format is that of Fig.~\ref{fig:coord} and the corresponding snapshots
of the impact are given in Fig.~\ref{fig:seq_21}
Left to right:  times 10, 20, 30 and 100$\tau$.  Top to bottom: 
potential energy density, fluid kinetic energy and particle kinetic energy.
The maximum heights in the left-hand frames are potential energy:
5.01$\epsilon$, fluid kinetic
energy: 3.92$\epsilon$ and particle kinetic energy: 61.2$\epsilon$.}
\label{fig:energy_21}
\end{center}
\end{figure}

\begin{figure}
\begin{center}
\includegraphics[width=0.9\linewidth]{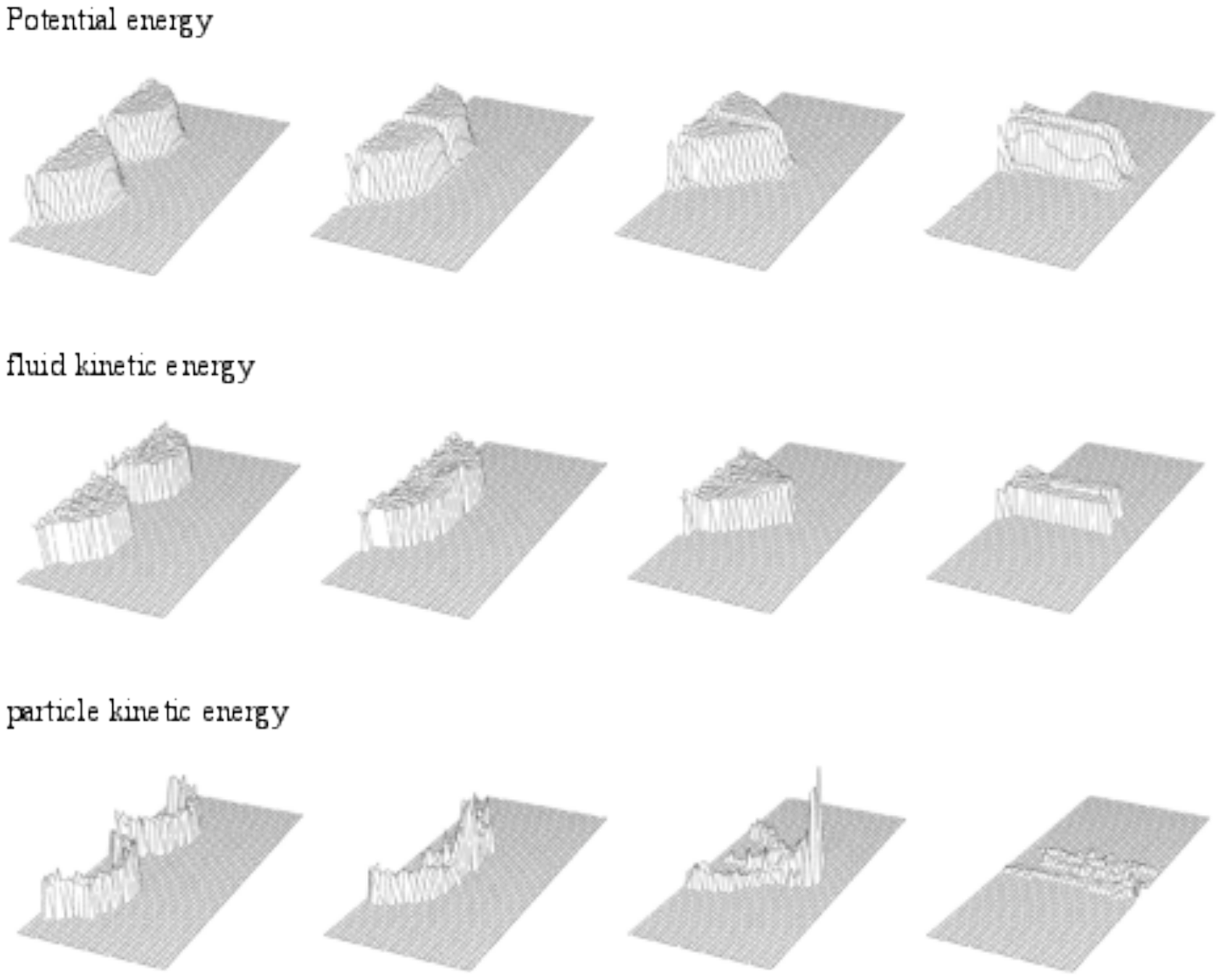}
\caption{Surface elevation plot of the two-dimensional density
fields arising in the collision of two ``opposite" 
liquid marble drops at velocity $2.0\sigma/\tau$.
The format is that of Fig.~\ref{fig:coord} and the corresponding snapshots
of the impact are given in Fig.~\ref{fig:seq_marb}
Left to right:  times 10, 30, 50 and 100$\tau$.  Top to bottom: 
potential energy density, fluid kinetic energy and particle kinetic energy.
The maximum heights in the left-hand frames are potential energy:
5.11$\epsilon$, fluid kinetic
energy: 2.06$\epsilon$ and particle kinetic energy: 24.4$\epsilon$.}
\label{fig:energy_marb}
\end{center}
\end{figure}

The field plots for the other drop impact cases, as well as those for a 
drop impact on a wetting surface, are similar in that the qualitative 
description in the previous paragraph is equally applicable. The one
distinction is that for smaller impact velocity the temperature and kinetic
energy increases are smaller, and for higher velocity the increases are 
larger.  However, interesting differences do appear in
the analogous plots for drop collisions, shown for two examples in
Fig.~\ref{fig:energy_21} and \ref{fig:energy_marb}
for both drops of the same material 
(21\% suspension at $u_0=3\sigma/\tau$), 
and opposite or non-attracting materials 
(marbles at relative velocity $u_0=2\sigma/\tau$),
respectively. Just as in the impact example just discussed, 
the trends and general behavior of the potential energy plot follows that of
the atomic density while the fluid kinetic energy follows the temperature, so
we focus on the three energy plots.  In collisions of drops of the same 
material, there is a smooth merger of the drops into a disk, with a
small transient rise in potential energy in the merger region but a 
persistent ridge in the fluid kinetic energy which propagates outward along the
mid-plane.  The total potential energy is increased by the collision but shows
little change subsequently: 
6.27$\epsilon$ per particle in equilibrium to 7.31 just after contact to
7.66 after 500$\tau$. (The energy is computed as the sum of negative LJ
energy and positive FENE energy, and the latter happens to have a larger
magnitude.) The fluid temperature is 0.8$\epsilon/k_B$ in
equilibrium and, similarly, stabilizes at 0.94 at the end of the simulation.
The particles all have an initial translational kinetic energy
but after colliding most come almost to rest except for those at the edge
of the merged drop which move outward to form a saucer before halting. 
In the case of opposite materials, the fluid energies dip at the mid-plane
reflecting the gap between the colliding drops, but the energy distribution
is otherwise similar to the previous case.  

\section{Mixing} \label{mix}

\begin{figure}
\begin{center}
\includegraphics[scale=0.8]{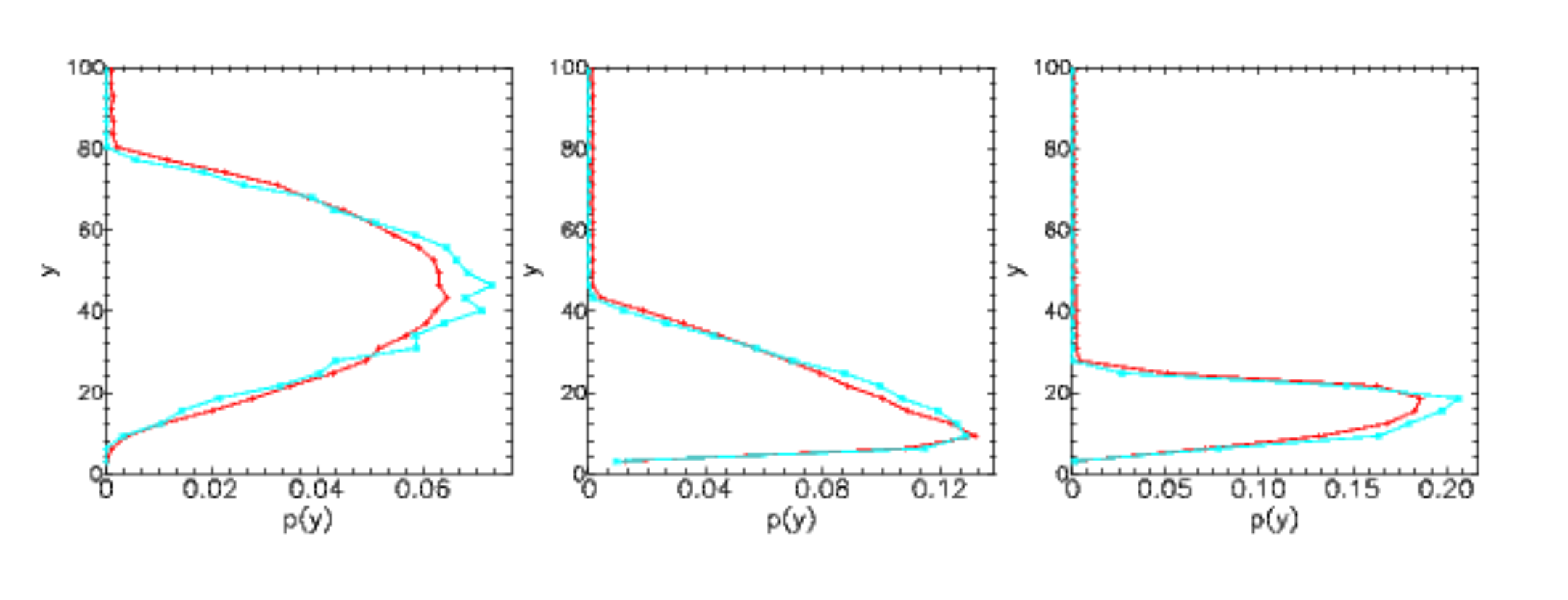}
\caption{Probability density of fluid (red +) and particles (cyan *)
as a function of height $y$ for the impact of a 21\% suspension drop on a
non-wetting wall at velocity $2\sigma/\tau$, at times (left to right) 10, 50
and 100$\tau$.}
\label{fig:imp_21_y}
\end{center}
\end{figure}

\begin{figure}
\begin{center}
\includegraphics[scale=0.75]{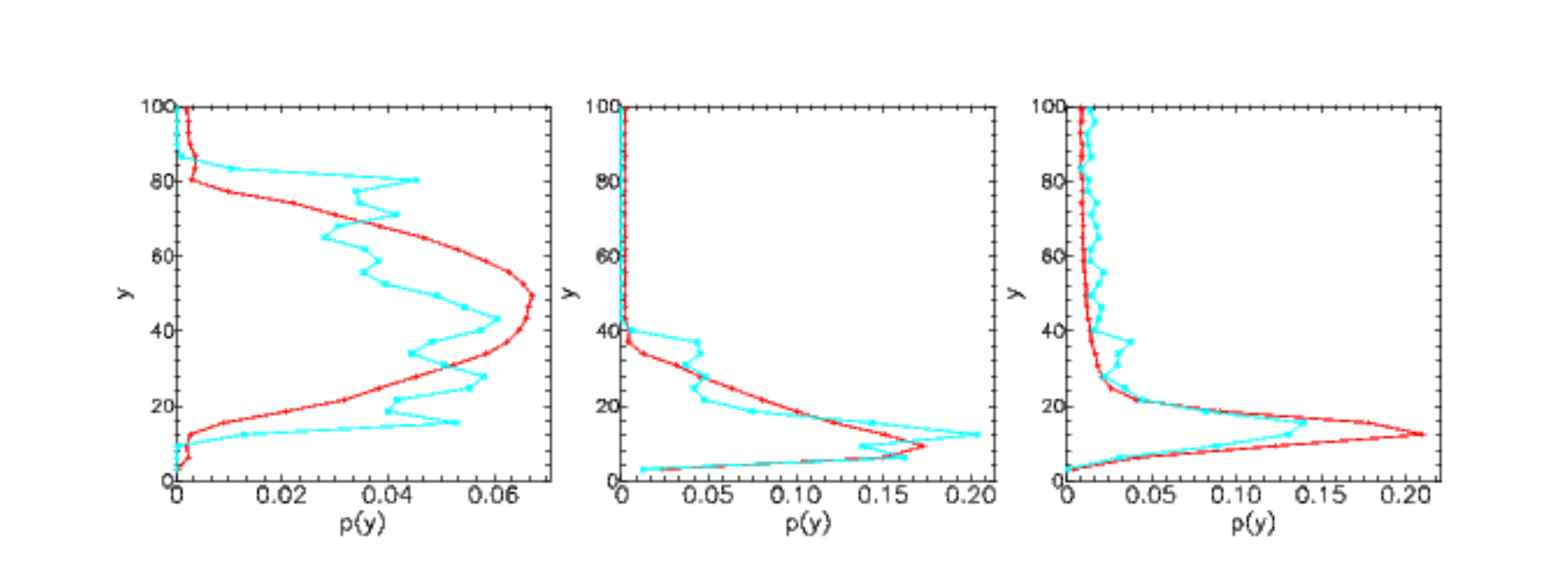}
\caption{Probability density of fluid (red +) and particles (cyan *)
as a function of height $y$ for the impact of a liquid marble drop on a
non-wetting wall at velocity $3\sigma/\tau$, at times (left to right) 10, 50
and 100$\tau$.}
\label{fig:imp_sh_y}
\end{center}
\end{figure}

The drops considered here have particles roughly uniformly distributed 
during the equilibration stage, but given the contrast in mass between the
fluid molecules and the solid particles, an impact may produce some relative
motion and inhomogeneity.  Our analysis is based on computing the
probability distribution for the various atomic species as a function of 
the coordinate $y$ along the impact axis.  We begin with the 21\% suspension
at low impact velocity $u_0=1.0\sigma/\tau$ on a non-wetting wall, shown in
Fig.~\ref{fig:imp_21_y}.  The particle and fluid distributions overlap well
just after impact, at maximum spreading and in the early stages of withdrawal
from the wall, indicating that uniform mixing is preserved.  Even at the higher
velocity $u_0=3\sigma/\tau$ where the drop disintegrates, fluid and particles
remain well mixed:  the peaks in the respective distributions always 
overlap, although there is a higher probability for vapor to occupy the 
space above the drop.  Impacts on a wetting wall likewise show little
separation of liquid and particles, although in this case there is always a
liquid layer adjacent to the wall.  In the denser 42\% suspension the
behavior is generally similar except for a
tendency for liquid to shoot out along the wall ahead of the particles at
early times, although the distributions overlap well afterwards.  Different 
behavior is seen in the liquid marble case in Fig.~\ref{fig:imp_sh_y}:
here the particle atoms are relatively weakly bound to the drop and are more
prone to separate.  The particles shoot out just after
impact and are preferentially located near the surface, while at later
times after bouncing off the surface they tend to lie above the liquid.
The figure illustrates this behavior for the more dramatic case 
$u_0=3\sigma/\tau$ but the same behavior is seen at lower velocities.
We conclude that as long as the particle atoms are strongly attracted to 
those of the liquid (in continuum terms, the liquid wets the particles),
uniform mixing is preserved in surface impacts.

\begin{figure}
\begin{center}
\includegraphics[scale=0.75]{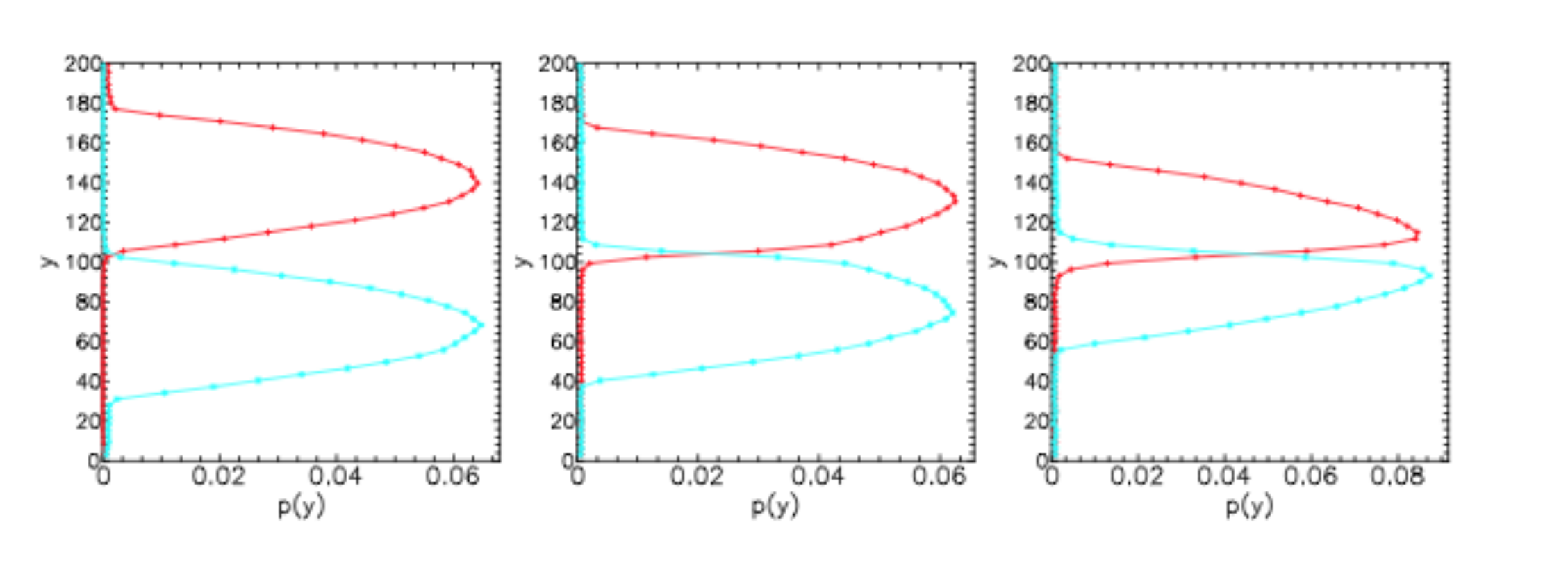}
\caption{Probability density of fluid initially in the upper (red +) and 
lower (cyan *) drops as a function of height $y$ for the coalescence of 
two identical liquid drops at times (left to right) 10, 200 and 500$\tau$.}
\label{fig:liq_coal}
\end{center}
\end{figure}

\begin{figure}
\begin{center}
\includegraphics[scale=0.65]{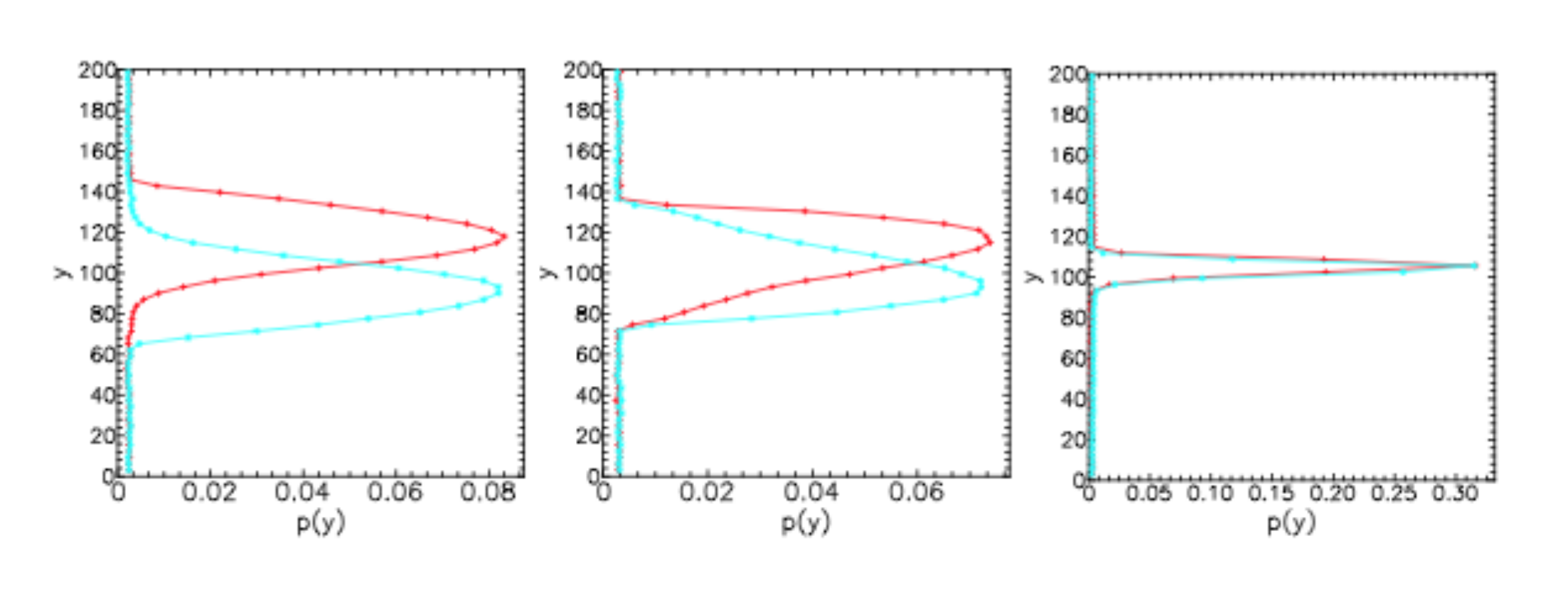}
\caption{Probability density of fluid initially in the upper (red +) and 
lower (cyan *) drops as a function of height $y$ for collision of 
two identical liquid drops at time 500$\tau$ for relative velocities
(left to right) 1, 2 and 3$\sigma/\tau$.}
\label{fig:same_liq_coll}
\end{center}
\end{figure}

A similar analysis of the relative motion of liquid and solid is relevant in 
drop collisions, and in addition there is the related question of 
whether there is
any enhancement in the rate or degree of mixing of the materials in the 
two colliding drops. In particular, we wish to contrast the behavior of
colliding drops to that of (quasi-static) coalescence where two drops are 
simply placed in close proximity and allowed to merge spontaneously.  
To this end, we have also simulated free coalescence by taking the initial
configuration of two liquid drops used in the collision calculations and
shifting the respective drop atomic positions towards each other so that the  
tails of their density profiles just overlap. The system was then allowed
to evolve freely, with no imposed relative velocities of the drops and no 
thermostat acting, and after a short transient period the two drops merged 
smoothly.  The evolution of the density pdf as a function of $y$ is shown in
Fig.~\ref{fig:liq_coal} for time up to 500$\tau$: the increasing overlap of 
the two original distributions 
reflects the degree of mixing of the drops.  The effect of collisions is to
increase the speed of this process, as shown in Fig.~\ref{fig:same_liq_coll}. 
When the drops collide with relative velocity 1.0$\sigma/\tau$, after
500$\tau$ the overlap region is twice as wide, even though the drop centers 
are slightly farther apart.  The latter property results from the fact that
the collision initially squeezes the drops close together, followed by a
partial recoil, and then a decaying oscillation at longer times.  At twice
the collision velocity, the oscillation is even stronger but at the same 
time 500$\tau$ the distribution overlap is much greater. At still higher
velocity, however, the collision produces a thin sheet which develops holes
and reaches the edge of the simulation box at 150$\tau$, preventing recoil.
Nonetheless, by this time the $y$-distributions overlap almost completely 
and the molecules of the original drops are very well mixed. The degree of
mixing thus increases monotonically with collision velocity.

\begin{figure}
\begin{center}
\includegraphics[width=0.5\linewidth]{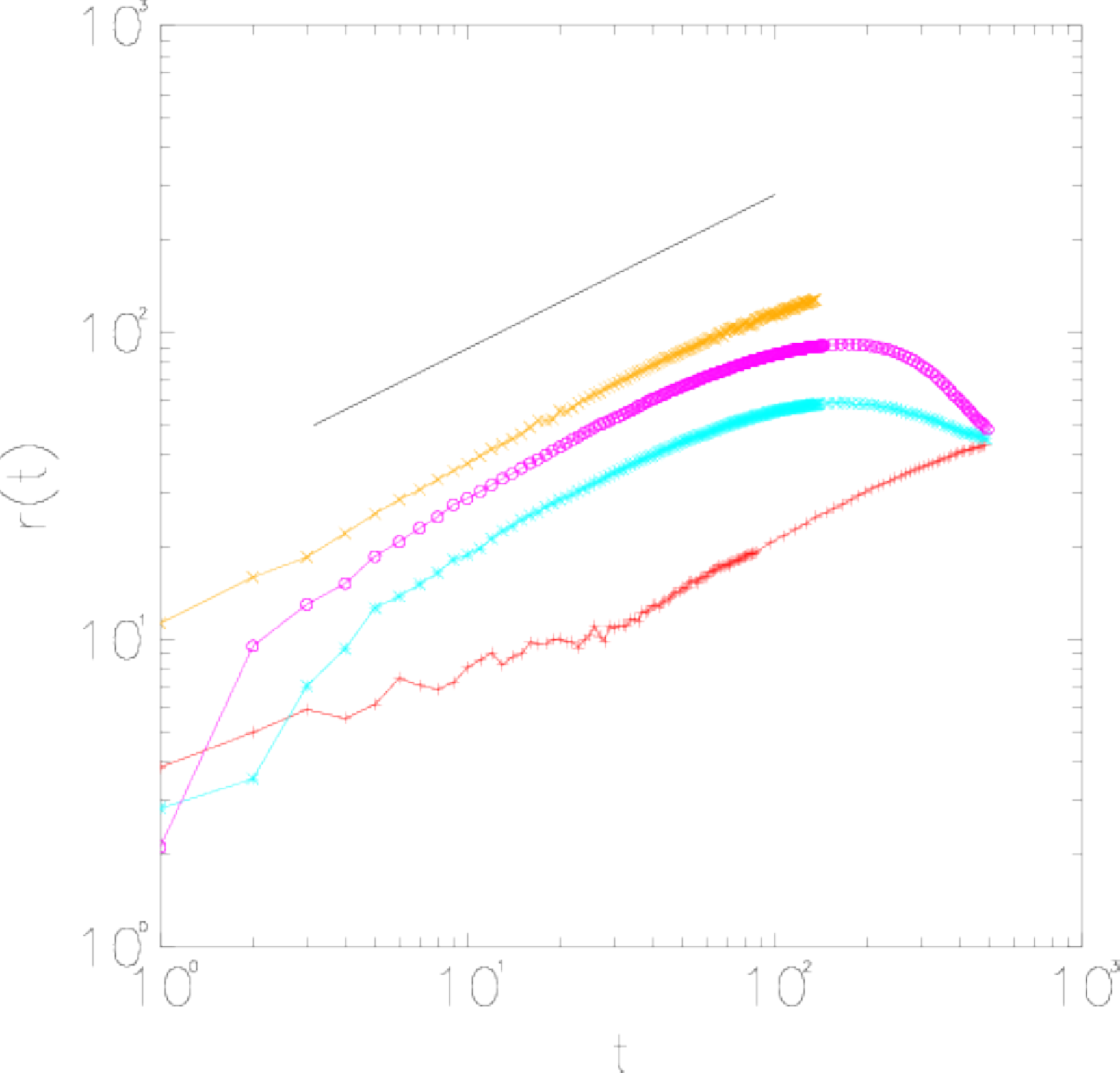}
\caption{Time dependence of the neck radius of two merging drops at various 
velocities, where $t=1$ is the earliest time at which a neck can be
identified. Bottom to top: neck radius for collisions at relative velocities 
0 (free coalescence, red), 1 (blue), 2 (magenta) and 3$\sigma/\tau$
(orange). The solid black line at the top shows $r(t)\sim t^{1/2}$ for 
comparison.}
\label{fig:neck}
\end{center}
\end{figure}

Although not the main focus of this paper, there has been extensive recent
interest in the time-dependence of the radius of the neck formed when two
drops coalesce, which provides an opportunity to compare experiment
and asymptotic analysis.  This information is available here as a
by-product of the coalescence simulation described above, and furthermore we
are able to track the growth of the neck radius for colliding drops as
a function of time.
The procedure is to examine the radially-averaged density field as a
function of $r$ and $y$, discussed above, and for each $y$ identify the 
liquid-vapor interface as the value of $r$ when the density equals half the
bulk value. Coalescence is taken to begin when this interface radius 
develops a non-zero minimum between the drop centers, and we plot the
minimum value of a function of the time $t$ after coalescence 
in Fig.~\ref{fig:neck}.  In all cases, 
ranging from free coalescence to violent collision at velocity
3$\sigma/\tau$, the neck radius is seen to vary as $t^{1/2}$ over one to
1-1/2 decades in time. Not
surprisingly, the growth rate itself (the prefactor) increases with impact
velocity, roughly linearly.  The variation is somewhat irregular at early 
times, due to thermal fluctuations in shape, until the
neck radius reaches a value of around 10$\sigma$ (neck diameter $\sim$ 20
atoms).
The curves turn over in the case of colliding drops, because there is
a residual oscillation after merger. We should emphasize that the behavior
shown here pertains to {\em nano}drops, where only about one decade of radius 
variation is available, and it is possible for other dynamical regimes to be
present when larger drops merge \cite{eggers,paulsen}. In fact, another 
MD simulation \cite{pothier} 
using a different choice of interatomic potentials found a crossover between
the expected linear growth law for the neck radius at early times and 
$t^{1/2}$ growth afterwards.  However, we note a recent lattice-Boltzmann
study \cite{gross}, ostensibly relevant to macroscopic drops, found the same 
result as seen here, a single square-root growth law for the neck radius.
 
\begin{figure}
\begin{center}
\includegraphics[width=0.95\linewidth]{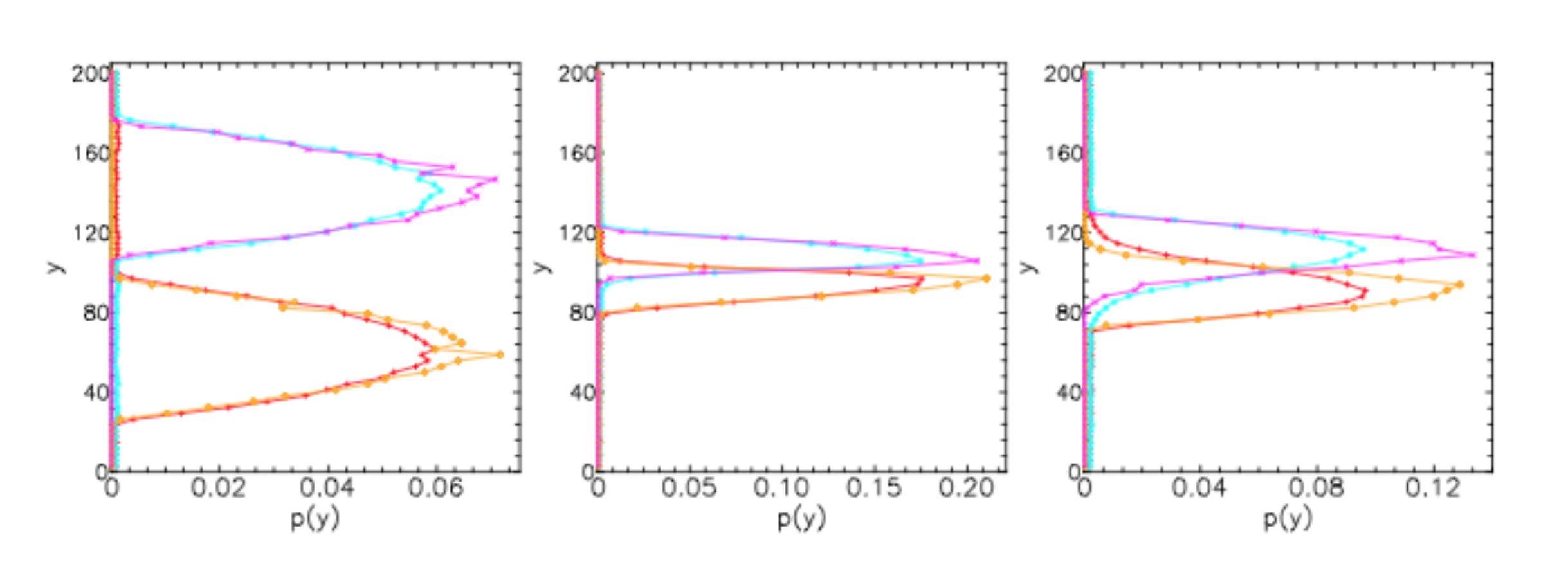}
\caption{Probability density of fluid and particles initially in the upper
drop (cyan * and magenta o, respectively) and likewise for the lower drop
(red + for fluid and orange x for particles) for the collision of two 21\%
suspension drops of the same material, at relative velocity $2\sigma/\tau$, 
as a function of height $y$, at times (left to right) 10, 100 and 500$\tau$.}
\label{fig:mix_21_v2}
\end{center}
\end{figure}

\begin{figure}
\begin{center}
\includegraphics[width=0.95\linewidth]{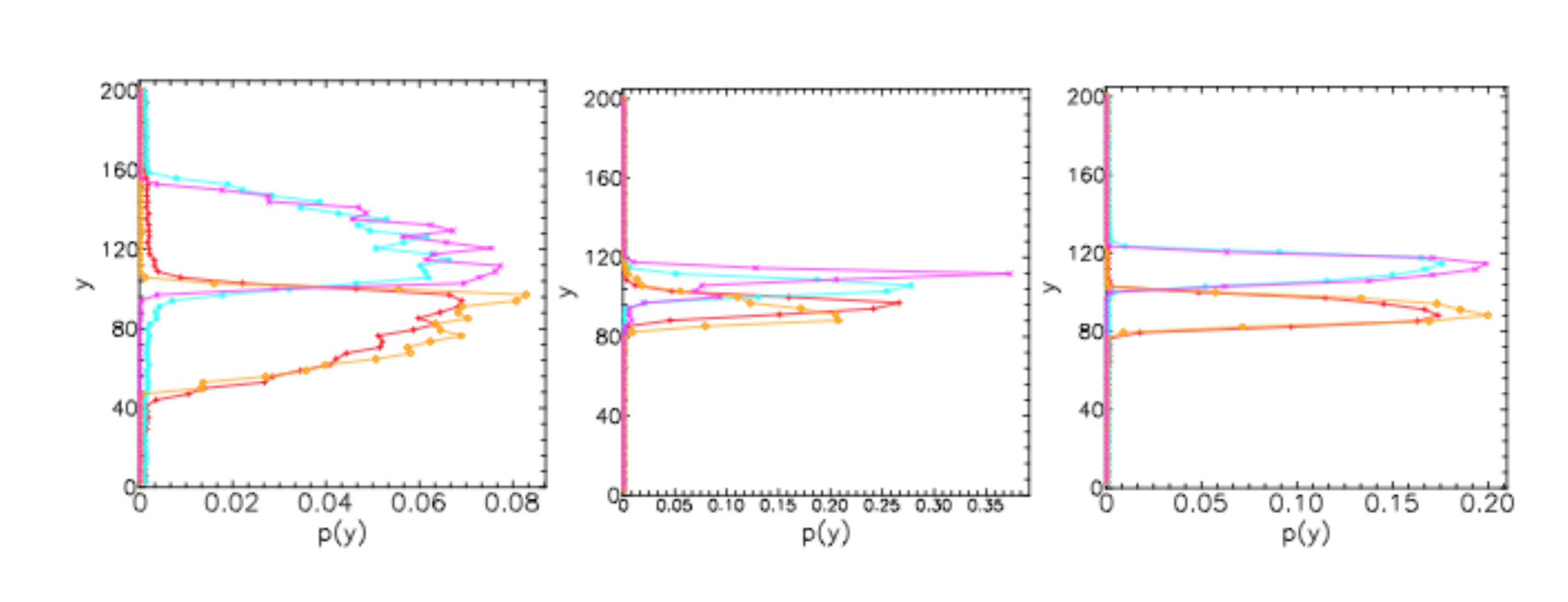}
\caption{Probability density of fluid and particles initially in the upper
drop (cyan * and magenta o, respectively) and likewise for the lower drop
(red + for fluid and orange x for particles) for 
Left: two 42\% suspension drops, Middle: liquid marble and Right:
``opposite'' 21\% suspension drops, all at relative velocity $2\sigma/\tau$
at time 100$\tau$.}
\label{fig:other_mix}
\end{center}
\end{figure}

Returning to the issue of mixing, the same analysis used for liquid drops 
is applied to the particle-laden drops, where there are
now four distributions, for the fluid and particles in each of the two initial
drops.  A typical example is show in Fig.~\ref{fig:mix_21_v2}, for 
the collision of identical 21\% suspension drops at relative velocity
2$\sigma/\tau$.  One sees the degree of mixing of the two drops increasing 
with time, but at a much slower rate than in the pure liquid case at this
velocity.  The
slower motion of the relatively massive particles presumably slows the mixing
process.  Another feature to note is that the particles tend to stay with
their original liquid, in the sense that for each drop the fluid and
particles distributions are nested.  The recoil of the drops after impact is
also evident in this plot. The same trends are observed at both lower and
higher collision velocities for this material, but different behavior is
seen in the other drops.  The 42\% suspension drops distort when they
collide, but after some weak initial mixing the $y$-distribution stabilize.
An example at velocity 2$\sigma/\tau$ is given in Fig.~\ref{fig:other_mix}a at
100$\tau$.  After this time there is little change in the spatial
distributions, and the drop's behavior is closer to that of a solid. As one
would expect, the
width of the mixed zone is broader and narrower at lower and higher
collision velocities, respectively, but always stabilizes after some
initial deformation.  In the case of the liquid marble drop, the particle
distribution no longer nests inside the fluid distribution, since the
liquid-solid atomic interaction is weaker and particles are expelled after
collision.  Nonetheless the presence of particles still slows the mixing of
the liquids, as indicated in Fig.~\ref{fig:other_mix}b, for velocity
2$\sigma/\tau$ at 100$\tau$.  A last variant case is the collision of
``nonwetting'' drops, whose respective atoms have only a repulsive
interaction.  Here the $y$-distributions of the atoms in the original drops 
remain separated after the collision, as illustrated in
Fig.~\ref{fig:other_mix}c for 21\% suspension drops at the same velocity 
and elapsed time.  

We wish to emphasize that while 
while we have provided detailed information for a only 
subset of the simulated cases in this section, the stated conclusions 
concerning the qualitative aspects of energy transfer and mixing are
generic, at least within the range of molecular systems studied here.

\section{Conclusions} 
\label{concl}

We have used MD simulations to study the surface impacts and collisions of
nanometer-sized suspension drops.  The range of impact velocities was chosen
to span the full range of phenomena from bouncing accompanied by distortion
and recovery to violent disintegration,
and collision simulations at the same relative velocities provided a direct
comparison of the two processes.  Only normal impacts and head-on
collisions were considered here, but other cases could be easily studied
using the same methods. The behavior of four types of drop were compared:
pure (diatomic) liquid, homogeneous suspensions of either ``low'' (21\%) or
``high'' (42\%) particle volume fractions and a ``liquid marble'' with a
surface coverage of roughly 40\%.  We have also carried out simulations for
drops with lower volume fractions, but these behave in a qualitatively similar 
manner to the 21\% suspension and marble cases presented here. Highly
concentrated suspensions exhibit distinctly different behavior, however,
which we have only begun to explore here.

A general conclusion is that the presence of nanoparticles in a drop, either
uniformly suspended or coating the surface, does not qualitatively alter 
its impact behavior unless the solid concentration is quite high.  In the
former case, corresponding to particles which are strongly wetted by the
liquid, changes in spreading radius and rate are seen but the behavior is
that of a liquid of higher viscosity and the fate of the drop is unchanged.  
In the latter case, where the solid-liquid interactions
are of intermediate strength and the particles reside at the surface, the
nanoparticles are driven by the bulk fluid motion at low impact speeds and
simply tend to fly off in more rapid impacts. For these drops the
distinctions between the bouncing, splashing, disintegration, coalescence
and stand-off regimes depend most critically on impact velocity.
The concentrated suspension drops tend to be more coherent in impact because
the nanoparticles here have a short-distance attraction to each other and in a
sense act to glue the drop together.  A limited analysis of the contrasting
case of no direct interparticle attraction shows the opposite behavior, with
the particles flying away from the impacting liquid region. More work is
needed to understand the relation between the concentrated suspension drop
simulations here and 
the flow of bulk suspension flows, where various non-linear rheological
behaviors are observed.
    
Some further specific conclusions on mixing and energy transfer are as follows.
\begin{enumerate}
\item The atomic density does not increase significantly in the contact
region as a result of the process. In a wetting impact or a collision of the
same materials, the density gap rapidly fills but does 
not rise substantially in height above the
bulk values before impact. In the opposite case of impact on a wetting
surface or a collision between atomically repelling materials, a density gap
of atomic thickness 
persists but there is no formation of a higher-density ridge of 
molecules or particles in the impact regions to either side.
\item In correspondence to the absence of an increase in density, there is 
no local enhancement in the potential energy.
\item The fluid kinetic energy and temperature both rise locally in the
contact region, to a degree that increases with the relative velocity.  
Much of the kinetic energy dissipates relatively quickly into the
thermostatted wall or the bulk of the drop(s)  but the temperature
enhancement persists over longer times.  In surface impact the drop
temperature eventually falls to that of the (thermostatted) wall but in
collisions the drop temperature can only equilibrate relatively slowly with 
the low-density surrounding vapor. 
\item Collisions are ``gentler'' than surface impacts, in the sense that 
significantly higher velocities are required to disintegrate a drop in
a collision ($u_0\sim 3-4\sigma/\tau$) than in an impact ($\sim 2\sigma/\tau$).
A rigid solid surface tends to reflect impinging molecules, causing
disruptive hard interactions with those in the interior of the drop, whereas in
a collision the molecules in the two drops are relatively free to rearrange
themselves locally without causing global changes in the configuration.
\item Drop collisions enhance the rate of mixing of identical drops of liquid,
but have only a modest effect in promoting fluid and particle mixing in 
the suspension
case.  When the materials are repelling, in the sense of having no atomic
attraction, a collision does not promote mixing unless it is so violent as
to disintegrate the drops. 
\end{enumerate}

The goal of this paper is to identify 
trends rather than to make quantitative predictions, since few experimental
results at this scale are available and furthermore it would be
difficult to precisely reproduce the simulated systems experimentally.
In a previous paper \cite{p2}, we were able to compare (mostly favorably) 
our results for low-velocity impact nanodrop 
simulation to experimental and theoretical scaling laws for 
continuum scale pure-liquid drop spreading. However, such
data is not available for the
particle-laden case and, in addition, in higher-velocity phenomena 
such as splashing the nanoscale simulations behave differently than
laboratory experiments.  A principal reason 
is that when the impact velocity is chosen to match the laboratory values of
Reynolds and Weber numbers where the transition to splashing is found, the
small drop size necessitates velocities so high that the Mach number becomes
appreciable and the post-impact temperature rises above the liquid-vapor
coexistence value.  A second issue is that the small size of the drops  
precludes the appearance of all of the regimes observed in larger systems,
which develop on longer space and time scales not present here.  The
obvious remedy -- bigger simulations -- is problematic because at least a
decade in size variation is required to pin down the scaling characteristics 
of any flow regime, and a decade increase in drop radius is a factor of 1000 
in the number of atoms.  MD codes are at best vary linearly in CPU time with
particle number, and furthermore bigger systems require longer physical
times to develop, so this approach is impractical at present. A more
promising means of connecting nano- and larger scale drop phenomena may be
the use of coarse-grained fluid mechanical models, such as discrete particle
dynamics, but in this case the atomic resolution is absent, and some effort
would be required to reliably connect such a model to molecular interactions 
and structure.

\newpage

\newpage

\newpage

\newpage

\newpage

\newpage

\newpage

\newpage
\begin{thebibliography}{99}
\bibitem{worth} A. M. Worthington, {\em A Study of Splashes} (Longmans,
Green, London, 1908).

\bibitem{rein} M. Rein, Phenomena of liquid drop impact on solid and liquid
surfaces, Fluid. Dyn. Res. {\bf 12}, 61-93 (1993).

\bibitem{yariv} A, L Yarin, Drop impact dynamics: splashing, spreading,
receding, bouncing, \ldots, Annu. Rev. Fluid Mech. {\bf 38}, 159-192 (2006).

\bibitem{lesser} M. B. Lesser and J. E. Field, The impact of compressible
liquids, Annu. Rev. Fluid Mech. {\bf 15}, 97-122 (1983).

\bibitem{marengo} M. Marengo, C. Antonini, I. V. Roisman, and C. Tropea, 
Drop collisions with simple and complex surfaces, Curr. Opin. Colloid Int.
Sci. {\bf 16}, 292-302 (2011).

\bibitem{crooks} R. Crooks and D. V. Boger, influence of fluid elasticity on
drops impacting on dry surfaces, J. Rheol {\bf 44},973-996 (2000).

\bibitem{german} G. German and V. Bertola, Impact of shear-thinning and
yield-stress drops on solid surfaces, J. Phys.--Condens. Mat. {\bf 21},
375111 (2009).

\bibitem{luu} L.-H. Luu and Y. Forterre, Drop impact of yield-stress fluids,
J. Fluid Mech. {\bf 632}, 301-327 (2009).

\bibitem{guemas} M. Gu\'emas, A. G. Marin and D. Lohse, Drop impact experiments
of non-Newtonian liquids on microstructured surfaces, Soft Matter {\bf 8},
10725-10731 (2012).

\bibitem{moon} J. H. Moon, J. B. Bong and S. H. Lee, Dynamic behaviorof
non-Newtonian droplets impinging on solid surfaces, Matls. Trans. {\bf 54},
260-265 (2013).

\bibitem{aussillous} P. Aussillous and D. Qu\'er\'e, Liquid marbles, Nature
{\bf 411}, 924-927 (2001); Properties of liquid
marbles, Proc. R. Soc. A {\bf 462}, 973-999 (2006).

\bibitem{nicolas} M. Nicolas, Spreading of a drop of neutrally buoyant
suspension, J. Fluid Mech. {\bf 545}, 271-280 (2005).

\bibitem{jaeger1} I. R. Peters, Q. Xu and H. M. Jaeger, Splashing onset in dense
suspension drops, Phys. Rev.Lett. {\bf 111}, 028301 (2013).

\bibitem{marston} J. O. Marston, M.M. Mansour and S. T. Thoroddsen Phys. Rev. E 
{\bf 88}, R010201 (2013).

\bibitem{jaeger2} L. A. Lubbers, Q. Xu, S. Wilken, W. W. Zhang and H. M.
Jaeger, Dense suspension splat:  monolayer spreading adn hole formation
after impact, Phys. Rev. Lett. {\bf 113}, 044502 (2014).

\bibitem{planchette} C. Planchette, A.-L. Biance, O. Pitois and E.
Lorenceau, Coalecence of armored interfaces under impact, Phys.
Fluids {\bf 25}, 042104 (2013).
 
\bibitem{p1} J. Koplik and R. Zhang, Nanodrop impact on solid surfaces, Phys.
Fluids {\bf 25}, 022003 (2013).

\bibitem{p2}  R. Zhang, S. Farokhirad, T. Lee and J. Koplik, Multiscale liquid
drop impact on wettable and textured surfaces, Phys. Fluids {\bf 26}, 082003
(2014).

\bibitem{ashgriz} N. Ashgriz and J. Y. Poo, Coalescence and separation in
binary collisions of liquid drops, J. Fluid Mech. {\bf 221}, 183-204 (1990).

\bibitem{qian} J. Qian and C. K. Law, J. Fluid Mech. {\bf 331}, 59-80 (1997).

\bibitem{eggers} J. Eggers, J. R. Lister and H. A. Stone, Coalescence of liquid
drops, J. Fluid Mech. {\bf 401}, 293-310 (1999).

\bibitem{roisman} I. V. Roisman, C. Planchette, E. Lorenceau and G. Brenn,
Binary collisions of drops of immiscible liquids, J. Fluid Mech. {\bf 690},
512-535 (2012).

\bibitem{paulsen} J. D. Paulsen, J. C. Burton, S. R. Nagel, S. Appathurai, M.
T. Harris and O. A. Basaran, The inexorable resistance of inertia
determines the initial regime of drop coalescence, Proc. Nat. Acad. Sci. USA
{\bf 109}, 6857-6861 (2012).

\bibitem{greenspan} D. Greenspan and L. F. Heath, Supercomputer simulation of
the modes of colliding microdrops of water, J. Phys. D {\bf 24}, 2121-2123
(1991).

\bibitem{koplik} J. Koplik and J. R. Banavar, Molecular structure of the
coalescence of liquid interfaces, Science {\bf 257}, 1664-1667 (1992).

\bibitem{svanberg} M. Svanberg, L. Ming, N. Markovi\'c and J. B. C.
Pettersson, Collision dynamics of large water clusters, J. Chem. Phys.
{\bf 108}, 5888-5897 (1998).

\bibitem{zhao} L. Zhao and P. Choi, Molecular dynamics simulation of the
coalescence of nanometer-sized waterdroplets in n-heptane, J. Chem. Phys. 
{\bf 120}, 1935-1942 (2004).

\bibitem{juang} R.-R. Juang, Y.-M. Lee, C.-H. Chang, J.-S. Wu, Y.-L. Hsu and
S.-W. Chau, Parallel molecular dynamics simulation of head-on collision of
two nanoscale droplets with low relative speed, J. Comput. Theor. Nanosci.
{\bf 6}, 46-53 (2009).

\bibitem{pothier} J.-C. Pothier and L. J. Lewis, Molecular dynamics study of
the viscous to inertial crossoever in nanodroplet coalescence, Phys. Rev. E
{\bf 85}, 115447 (2012).

\bibitem{kompinski} E. Kompinski and E. Sher, Experimental comparisons
between droplet-droplet collision and single-droplet impacts on a solid
surface, Atomization and Sprays {\bf 19}, 409-423 (2009).

\bibitem{gross} M. Gross, I. Steinbach, D. Raabe and F. Varnik, Viscous
coalescence of droplets: A lattice Boltzmann study, Phys. Fluids {\bf 25},
052101 (2013).

\bibitem{nagel} L.Xu, W.W. Zhang and S. R.Nagel, Drop splashing on a dry smooth
surface, Phys. Rev.Lett. {\bf 94}, 184505 (2005).

\bibitem{at} M. P. Allen and D. J. Tildesley, {\sl Computer Simulation
of Liquids} (Oxford, Clarendon Press, 1987).

\end{thebibliography}
\end{document}